\documentclass[12pt,journal,onecolumn]{IEEEtran}

\usepackage{amssymb,amsmath,amsfonts,bm,epsfig,graphicx,theorem,latexsym}
\usepackage{rotating,setspace,latexsym,epsf,color,subfigure}
\usepackage{cite,times,authblk,epstopdf}
\usepackage{multicol}%
\usepackage{verbatim}%
\usepackage{enumerate}
\usepackage{eucal}
\usepackage{caption}
\usepackage{longtable}
\usepackage{stmaryrd,bbm,subfigure,units}

\newtheorem{theorem}{Theorem}

\newtheorem{corollary}{Corollary}

\newtheorem{lemma}{Lemma}

\newtheorem{remark}{Remark}

\newenvironment{Proof}[1]{\medskip\par\noindent{\bf Proof:\,}\,#1}{{\mbox{\,$\blacksquare$}\medskip\par}}

\makeatletter

\allowdisplaybreaks[1]

\newcommand{\bX}{\bold{X}}
\newcommand{\bXh}{\hat{\bold{X}}}
\newcommand{\bx}{\bold{x}}
\newcommand{\bxh}{\hat{\bold{x}}}
\newcommand{\bY}{\bold{Y}}
\newcommand{\by}{\bold{y}}
\newcommand{\bZ}{\bold{Z}}
\newcommand{\tbZ}{\tilde{\bold{Z}}}
\newcommand{\bz}{\bold{z}}
\newcommand{\bV}{\bold{V}}
\newcommand{\bv}{\bold{v}}
\newcommand{\bU}{\bold{U}}
\newcommand{\bu}{\bold{u}}
\newcommand{\tM}{\tilde{M}}
\newcommand{\tC}{\tilde{C}}
\newcommand{\tP}{\tilde{P}}

\newcommand{\setA}{\mathcal{A}}
\newcommand{\setB}{\mathcal{B}}
\newcommand{\setC}{\mathcal{C}}
\newcommand{\setD}{\mathcal{D}}
\newcommand{\setF}{\mathcal{F}}
\newcommand{\setG}{\mathcal{G}}
\newcommand{\setK}{\mathcal{K}}
\newcommand{\setM}{\mathcal{M}}
\newcommand{\setN}{\mathcal{N}}
\newcommand{\setP}{\mathcal{P}}
\newcommand{\setU}{\mathcal{U}}

\newcommand{\setS}{\mathcal{S}}
\newcommand{\setT}{\mathcal{T}}
\newcommand{\setV}{\mathcal{V}}
\newcommand{\setX}{\mathcal{X}}
\newcommand{\setY}{\mathcal{Y}}
\newcommand{\setZ}{\mathcal{Z}}

\newcommand{\E}{{\mathbb{E}}}
\newcommand{\V}{{\mathbb{V}}}
\newcommand{\D}{{\mathbb{D}}}
\newcommand{\Prob}{{\mathbb{P}}}
\newcommand{\Var}{{\mathbb{V}{\rm{ar}}}}
\newcommand{\limitn}{{\underset{n\rightarrow\infty}\lim}}

\title{A New Wiretap Channel Model and its Strong Secrecy Capacity
\thanks{This paper was presented in part at the $2016$ IEEE International Symposium on Information Theory. This work was supported by NSF Grants CCF 13-19338 and CNS 13-14719. }}
\author{Mohamed Nafea}
\author{Aylin Yener}
\affil{\normalsize Wireless Communications and Networking Laboratory (WCAN)\\
Electrical Engineering Department\\
The Pennsylvania State University, University Park, PA 16802.\\
\em mnafea@psu.edu \qquad yener@engr.psu.edu}
\begin{document}

\IEEEoverridecommandlockouts

\maketitle
\vspace{-0.3cm} 

\begin{abstract}
In this paper, a new wiretap channel model is proposed, where the legitimate transmitter and receiver communicate over a discrete memoryless channel. The wiretapper has perfect access to a fixed-length subset of the transmitted codeword symbols of her choosing. Additionally, she observes the remainder of the transmitted symbols through a discrete memoryless channel. This new model subsumes the classical wiretap channel and wiretap channel II with noisy main channel as its special cases. The strong secrecy capacity of the proposed channel model is identified. Achievability is established by solving a dual secret key agreement problem in the source model, and converting the solution to the original channel model using probability distribution approximation arguments. In the dual problem, a source encoder and decoder, who observe random sequences independent and identically distributed according to the input and output distributions of the legitimate channel in the original problem, communicate a confidential key over a public error-free channel using a single forward transmission, in the presence of a compound wiretapping source who has perfect access to the public discussion. The security of the key is guaranteed for the exponentially many possibilities of the subset chosen at wiretapper by deriving a lemma which provides a {\it{doubly-exponential}} convergence rate for the probability that, for a fixed choice of the subset, the key is uniform and independent from the public discussion and the wiretapping source's observation. The converse is derived by using Sanov's theorem to upper bound the secrecy capacity of the new wiretap channel model by the secrecy capacity when the tapped subset is randomly chosen by nature. 
\end{abstract}

\section{Introduction}\label{Int}
Wyner's wiretap channel models a legitimate transmitter and a receiver communicating over a discrete memoryless channel (DMC), referred to as the main channel, in the presence of a passive wiretapper who {\it{only listens}} to the transmitted signal through a cascaded second DMC, referred to as the wiretapper channel \cite{WTCWyner}. Subsequently, reference \cite{CK} has generalized Wyner's wiretap channel model to a general, not necessarily degraded, discrete memoryless wiretap channel. Later, Ozarow and Wyner, in reference \cite{WTCII_Wyner}, have introduced the wiretap channel II model, which considers a noiseless main channel and a binary erasure channel to the wiretapper, where the wiretapper is able to select the positions of erasures. Interestingly, using random partitioning and combinatorial arguments, reference \cite{WTCII_Wyner} has showed that the secrecy capacity for this channel is equal to that if the wiretapper is a passive observer unable to choose the positions of the erasures, thus demonstrating the ability of coding to neutralize this more powerful wiretapper. 

While considerable research on code design for secure communication followed the randomized coset coding of \cite{WTCII_Wyner}, see for example \cite{thangaraj2007applications,liu2007secure,aggarwal2009wiretap}, the idea of the wiretap channel II remained linked to the assumption of a noiseless main channel for several decades, mainly due to technical challenges in generalizing the model outside of this special model. Yet, the notion of providing the wiretapper with this additional capability of choosing what to observe is appealing and represents a positive step towards providing confidentiality guarantees in stronger attack models. Towards this end, reference \cite{nafea2015wiretap} introduced a discrete memoryless (noisy) main channel to the wiretap channel II model, and derived outer and inner bounds for the capacity-equivocation region of the model, where the proposed achievability scheme is optimal for the special case of the maximizing input distribution being uniform. More recently, reference \cite{goldfeld2015semantic} found the secrecy capacity of this model, showing that, once again, the secrecy capacity is equal to that of the case when the wiretapper channel is replaced with a discrete memoryless erasure channel.

This work goes one step further and introduces a {\it{new wiretap channel model}} with a discrete memoryless main channel and a wiretapper who observes a subset of the transmitted codeword symbols of her choosing perfectly, as well as observing the remaining symbols through a second DMC. This general model includes as special cases both the classical wiretap channel in \cite{CK} by setting the subset size to zero, and the wiretap channel II with a noisy main channel in \cite{nafea2015wiretap} by setting the wiretapper's DMC to an erasure channel with erasure probability one. We characterize the {\it{strong}} secrecy capacity for the proposed wiretap channel model, quantifying precisely the cost in secrecy capacity due to the additional capability at the wiretapper, with respect to the previous wiretap models.

We first present the achievability. Recent independent work \cite{goldfeld2015semantic} has provided an achievability proof  for the channel model considered in \cite{nafea2015wiretap} using a stronger version of Wyner's soft covering lemma \cite{wyner1975cmmon}. Extending the achievability proof in \cite{goldfeld2015semantic} to the new wiretap channel model is challenging due to the additional noisy observations at the wiretapper, since the aforementioned lemma entails approximating the distribution of the wiretapper channel output with an independent and identically distributed (i.i.d.) $n$-letter distribution. Instead, we establish the achievability by a framework similar to the output statistics of random binning framework in \cite{yassaee2014achievability}. In particular, we solve a dual secret key agreement problem in the source model sense \cite{maurer1993secret,ahlswede1993common}, and infer the design for the encoder and decoder of the original channel model from the solution of the dual problem. The difference between our achievability proof and the framework presented in \cite{yassaee2014achievability} is that we measure the statistical dependence between the transmitted message and the wiretapper's observation in terms of the Kullback-Leibler (K-L) divergence instead of total variation distance, which requires establishing a convergence result, with a rate strictly faster than $\frac{1}{n}$, for the probability that the two induced distributions from the original and the dual models are close in the total variation distance sense. In addition, in the source model, we guarantee the secrecy of the confidential key for the exponentially many possibilities of the subset chosen at the wiretapper by deriving a {\it{one-shot}} result which provides a {\it{doubly-exponential convergence rate}} for the probability that the key is uniform and independent from the wiretapper's observation. To summarize, the main advantage of working with the appropriate dual source coding problem is that it renders the analysis of the scenario at hand tractable. 

The converse is derived by identifying a channel model whose secrecy capacity is identical to that of the proposed channel model, and is easier to establish the converse of. This is done by means of upper bounding its secrecy capacity with that of a discrete memoryless channel whose secrecy capacity is tractable.

The remainder of the paper is organized as follows. Section \ref{ChannelModel} describes the new wiretap channel model. Section \ref{MainResult} provides the main result of the paper, i.e., the strong secrecy capacity for the new wiretap channel. Sections \ref{Achievability} and \ref{Converse} provide the achievability and converse proofs. Section \ref{Con} concludes the paper. The proofs for the supporting lemmas are provided in the Appendices.

\section{Channel Model and Definitions}{\label{ChannelModel}}
We first remark the notation we use throughout the paper. Vectors are denoted by bold lower-case super-scripted letters while their components are denoted by lower-case sub-scripted letters. A similar convention  but with upper-case letters is used for random vectors and their components. Vector superscripts are omitted when dimensions are clear from the context. We use $\mathbbm{1}\{\setA\}$ to denote the indicator function of the event $\setA$. For $a,b\in \mathbb{R}$, $\llbracket a,b\rrbracket$ denotes the set of integers $\{i\in\mathbb{N}:a\leq i\leq b\}$. For $S\subseteq\mathbb{N}$, ${\bf{X}}_S$ denotes the sequence $\{X_i\}_{i\in S}$. We use upper-case letters to denote random probability distributions, e.g., $P_X$, and lower-case letters to denote deterministic probability distributions, e.g., $p_X$. We use $p_X^U$ to denote a uniform distribution over the random variable $X$. The argument of the probability distribution is omitted when it is clear from its subscript. $\V(p_X,q_X)$ and $\D(p_X||q_X)$ denote the total variation distance and the Kullback-Leibler (K-L) divergence between the probability distributions $p_X$ and $q_X$.

\begin{figure}
    \centering
	\includegraphics[scale=0.8]{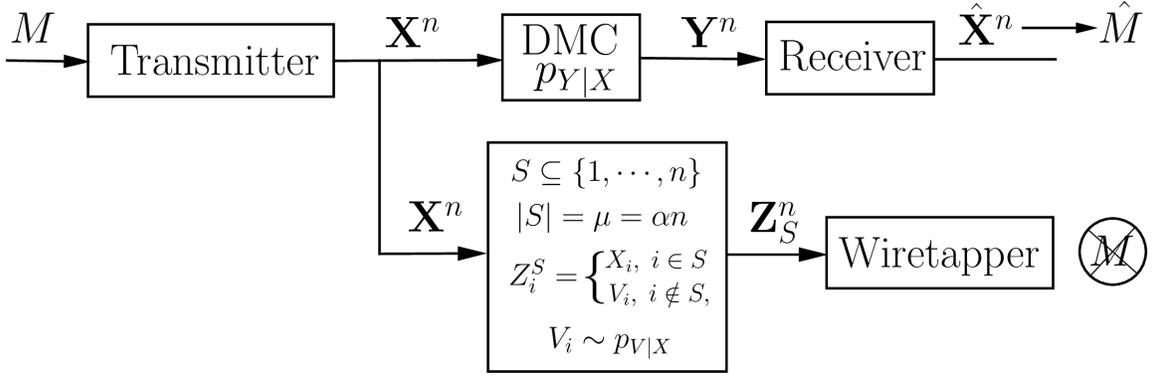}
	\caption{A new wiretap channel model.}
	\label{fig:sysmodel}
\end{figure}

We consider the channel model illustrated in Figure \ref{fig:sysmodel}. The main channel $\left\{\setX,\setY,p_{Y|X}\right\}$ is a discrete memoryless channel (DMC) which consists of a finite input alphabet $\setX$, a finite output alphabet $\setY$, and a transition probability $p_{Y|X}.$ The transmitter wishes to transmit a message $M$, uniformly distributed over $\setM=\llbracket 1,2^{nR_s}\rrbracket$, to the legitimate receiver reliably, and to keep the message secret from the wiretapper. To do so, the transmitter maps the message $M$ to the transmitted codeword $\bX^n\in\setX^n$ using a stochastic encoder. The legitimate receiver observes $\bY^n\in\setY^n$ and maps its observation to the estimate $\hat{M}$ of the message $M$. The wiretapper chooses a subset $S\in\setS$ where the set $\setS$ is defined as 
\begin{align}
\label{eq:setS}
\setS=\left\{S: S\subseteq \llbracket 1,n\rrbracket,\;\;|S|=\mu\leq n,\;\alpha=\frac{\mu}{n}\right\}.
\end{align}
Then, the wiretapper observes the sequence $\bZ_S^n=[Z_1^S,Z_2^S,\cdots,Z_n^S]\in {\mathcal{Z}^n}$, with 
\begin{align}
\label{eq:Z_S_n}
Z_i^S=\begin{cases}
X_i,\quad i\in S\\
V_i,\quad\;\text{otherwise},
\end{cases}
\end{align}   
where $\bV^n=[V_1,V_2,\cdots,V_n]\in\mathcal{V}^n$ is the output of the DMC $p_{V|X}$ when $\bX^n$ is the input, and the alphabet $\setZ$ is given by $\mathcal{Z}=\left\{\mathcal{X}\cup\mathcal{V}\right\}$.

An $(n,2^{nR_s})$ code $\setC_n$ for the  channel model in Figure \ref{fig:sysmodel} consists of 
\begin{enumerate}[(i)]
\item  the message set $\setM=\llbracket 1,2^{nR_s}\rrbracket$,
\item the stochastic encoder $P_{\bX^n|M, \setC_n}$ at the transmitter, and 
\item the decoder at legitimate receiver. 
\end{enumerate} 
We consider the strong secrecy constraint at the wiretapper \cite{csiszar1996almost,maurer2000information}. Rate $R_s$ is an achievable strong secrecy rate if there exists a sequence of $(n,2^{nR_s})$ channel codes, $\{\mathcal{C}_n\}_{n\geq 1}$, such that
\begin{align}
\label{eq:reliab_const}
&\underset{n\rightarrow\infty}\lim\Prob\left(\hat{M}\neq M|\mathcal{C}_n\right)=0\qquad\textbf{ Reliability},\\
\label{eq:sec_const}
\text{and}\quad&\underset{n\rightarrow\infty}\lim\underset{S\in\setS}\max\;I\left(M;\bZ_S^n|\mathcal{C}_n\right)=0 \qquad \textbf{Strong Secrecy},
\end{align}
where $\setS$ is defined as in (\ref{eq:setS}). The strong secrecy capacity, $\mathsf{C}_s$, is the supremum of all achievable strong secrecy rates.

Finally, we will be using the following two measures extensively in the sequel. The total variation distance between two probability distributions $p_X$ and $q_X$, defined on the same probability space, is given by 
\begin{align}
\label{eq:V_2}
\V(p_X,q_X)&=\frac{1}{2}\sum_{x\in\setX}|p(x)-q(x)|=\sum_{x\in\setX:\;p(x)>q(x)}[p(x)-q(x)]. 
\end{align}
The Kullback-Leibler divergence, or relative entropy, between the two distributions $p_X$ and $q_X$, defined on the same probability space, is given by 
\begin{align}
\label{eq:KL_1}
\D(p_X||q_X)&=\sum_{x\in\setX}p_X(x)\log \frac{p_X(x)}{q_X(x)}.
\end{align}

\section{Main Result}\label{MainResult}
The main result of this paper is stated in the following theorem. 
\begin{theorem}\label{thm:Thm1}
For $0\leq\alpha\leq 1$, the strong secrecy capacity of the new wiretap channel model in Figure \ref{fig:sysmodel} is given by 
\begin{align}
\label{eq:Thm1} 
\mathsf{C}_s(\alpha)=\underset{U-X-YV}\max\left[I(U;Y)-I(U;V)-\alpha I(U;X|V)\right]^+,
\end{align}
where the maximization is over all the distributions $p_{UX}$ which satisfy the Markov chain $U-X-YV$, and the cardinality of $U$ can be restricted as $|\setU|\leq |\setX|$. 
\end{theorem}
\begin{Proof}
The achievability and converse proofs for Theorem \ref{thm:Thm1} are provided in Sections \ref{Achievability} and \ref{Converse}, respectively.
\end{Proof}

\begin{remark}
{\emph{An equivalent characterization for the strong secrecy capacity of the new wiretap channel model is given by
\begin{align}
\label{eq:remark_2_1} 
\mathsf{C}_s(\alpha)=\underset{U-X-YV}\max\left[I(U;Y)-\alpha I(U;X)-(1-\alpha)I(U;V)\right]^+,
\end{align}
since $I(U;X|V)$ in (\ref{eq:Thm1}) can be written as 
\begin{align}
\label{eq:remark_2_2} 
I&(U;X|V)=H(U|V)-H(U|X)\\
\label{eq:remark_2_3} 
&=H(U)-I(U;V)-H(U|X)=I(U;X)-I(U;V),
\end{align}
where (\ref{eq:remark_2_2}) follows from the Markov chain $U-X-V$.}}
\end{remark}

\begin{corollary}
{\emph{By setting the tapped subset by the wiretapper, $S$, to the null set, or equivalently $\alpha=0$, the secrecy capacity in (\ref{eq:Thm1}) is equal to the secrecy capacity of the discrete memoryless wiretap channel in \cite[Corollary 2]{CK}, i.e., 
\begin{align}
\label{eq:remark_3_1}
\mathsf{C}_s(0)=\underset{U-X-YV}\max\left[I(U;Y)-I(U;V)\right]^+.
\end{align}}}
\end{corollary}

\begin{remark}
{\emph{Comparing (\ref{eq:Thm1}) and (\ref{eq:remark_3_1}), we observe that the secrecy cost, with respect to the classical wiretap channel, of the additional capability of the wiretapper to choose a subset of size $\alpha n$ of the codewords to access perfectly, is equal to $\alpha I(U;X|V)$.}} 
\end{remark}

\begin{corollary}
{\emph{By setting the wiretapper's DMC through which she observes the $(1-\alpha)n$ symbols she does not choose, $p_{V|X}$, to be an erasure channel with erasure probability one, the secrecy capacity in (\ref{eq:Thm1}) is equal to the secrecy capacity of the wiretap channel II with a noisy main channel in \cite[Theorem 2]{goldfeld2015semantic}, i.e., 
\begin{align}
\label{eq:remark_4_1} 
\mathsf{C}_s(\alpha)=\underset{U-X-Y}\max\left[I(U;Y)-\alpha I(U;X)\right]^+.
\end{align}}}
\end{corollary}

\begin{remark}
{\emph{Comparing (\ref{eq:remark_2_1}) and (\ref{eq:remark_4_1}), the secrecy cost, with respect to the wiretap channel II with a noisy main channel, of the additional capability of the wiretapper of observing $(1-\alpha)$ fraction of the codeword through the DMC $p_{V|X}$, is equal to $(1-\alpha) I(U;V)$.}}
\end{remark}

\section{Achievability}{\label{Achievability}}
We establish the achievability for Theorem \ref{thm:Thm1} using an indirect approach as in \cite{renes2011noisy,cuff2013distributed,yassaee2014achievability}. We define a {\it{dual}} secret key agreement problem in the source model which introduces a set of random variables similar to the set of variables introduced by the original problem. The alphabets of the random variables in the original and dual problems are identical. A subset of the random variables in the dual problem are considered to have distributions identical to the distributions of the corresponding variables in the original problem. Yet, the distribution of the random variables can differ from those of the original problem due to the different dynamics in the original and dual problems. The main trick is to search for conditions such that the joint distributions of the random variables in the two problems are almost identical in the total variation distance sense. This enables converting the solution, i.e., finding an encoder and decoder which satisfy certain reliability and secrecy conditions, for the dual problem, which is more tractable, to a solution of the original problem. Duality here is an {\it{operational}} duality \cite{gupta2011operational} in which the solution for the dual problem is converted to a solution for the original problem.

We first prove the achievability for the case $U=X$. We fix the input distribution $p_X$ and define two protocols; each of these protocols introduces a set of random variables and random vectors and induces a joint distribution over them. The first protocol, protocol A, describes a dual secret key agreement problem in which a source encoder and decoder observe random sequences independently and identically distributed (i.i.d.) according to the input and output distributions of the original channel model. The source encoder and decoder intend to communicate a confidential key via transmitting a public message over an error-free channel, in the presence of a {\it{compound}} wiretapping source who has perfect access to the public message and observes another random sequence whose distribution belongs to a finite class of distributions, with no prior distribution over the class. The second protocol, protocol B, describes the original channel model in Figure \ref{fig:sysmodel}, with the addition of assuming a common randomness that is available at all terminals. 

We first derive rate conditions for protocol A such that its induced distribution is close in the total variation distance sense to the induced distribution from protocol B. Then, we derive rate conditions for protocol A such that (i) the communication of the key is reliable, and that (ii) the probability of the key being uniform and independent from the public message and the wiretapping source observation converges doubly exponentially to one with the block-length. Next, we use the closeness of the two induced distributions from the two protocols to show that, under the same rate conditions for protocol B, properties (i) and (ii) hold for protocol B as well. Finally, we eliminate the assumed common randomness from protocol B by conditioning on a certain instance of the common randomness. Property (ii) for protocol B after removing the common randomness results in an achievable strong secrecy rate for the original channel model. In the following, we describe the two protocols in detail.

\begin{figure}
    \centering
	\includegraphics[scale=.85]{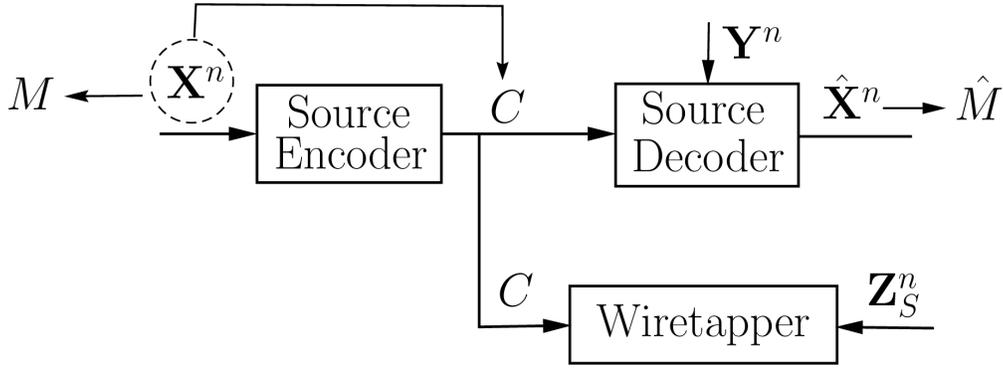}
	\caption{Protocol A: Secret key agreement in the source model.}
	\label{fig:Protocol A}
\end{figure}

{\it{Protocol A (Secret key agreement in source model):}} The protocol is illustrated in Figure \ref{fig:Protocol A}. The random vectors $\bX^n,\bY^n$ are i.i.d. according to $p_{XY}=p_X p_{Y|X}$, where $p_{Y|X}$ is the transition probability of the main channel in Figure  \ref{fig:sysmodel}. The source encoder observes the sequence $\bX^n$ and randomly assigns (bins) it into the two bin indices $M=\setB_{1,n}(\bX^n)$ and $C=\setB_{2,n}(\bX^n)$, where $\setB_{1,n}$ and $\setB_{2,n}$ are uniformly distributed over $\llbracket 1,2^{nR_s} \rrbracket$ and $\llbracket 1,2^{n\tilde{R}_s}\rrbracket$, respectively. That is, each $\bx^n\in\setX^n$ is randomly and independently assigned to the indices $m\in\llbracket 1,2^{nR_s} \rrbracket$ and $c\in\llbracket 1,2^{n\tilde{R}_s} \rrbracket$. The bin index $C$ represents the public message which is transmitted over a noiseless channel to the decoder and perfectly accessed by the wiretapper. The bin index $M$ represents the confidential key to be generated at the encoder and reconstructed at the decoder. The source decoder observes $C$ and the i.i.d. sequence $\bY^n$, and outputs the estimate $\bXh^n$ of $\bX^n$, which in turn generates the estimate $\hat{M}$ of $M$. For any $S\in\setS$, where $\setS$ is defined as in (\ref{eq:setS}), the wiretapper source node observes $C$ and the sequence $\bZ_S^n$ in (\ref{eq:Z_S_n}). The subset $S$ is selected by the wiretapper and her selection is unknown to the legitimate parties. Thus, the wiretapper can be represented as a compound source ${\bZ_S^n}\triangleq\left\{\mathcal{Z},{p_{\bZ_S^n},S\in\setS}\right\}$ whose distribution is only known to belong to the finite class $\{p_{\bZ_S^n}\}_{S\in\setS}$ with no prior distribution over the class, with $|\setS|=\binom{n}{\alpha n}\leq 2^{n}$. For $S\in\setS$, the induced joint distribution for this protocol is 
\begin{align}
\tP_{MC\bX\bY\bZ_S\bXh}(m,c,&\bx,\by,\bz,\bxh)=p_{\bX\bY\bZ_S}(\bx,\by,\bz) \tP_{MC|\bX}(m,c|\bx) \tP_{\bXh|\bY C}(\bxh|\by,c)\\
\label{eq:P_tilde_1}
&=p_{\bX\bY\bZ_S}(\bx,\by,\bz)\mathbbm{1}\{\setB_{1,n}(\bX)=M\}\mathbbm{1}\{\setB_{2,n}(\bX)=C\}\tP_{\bXh|\bY C}(\bxh|\by,c)\\
\label{eq:P_tilde_2}
&=\tP_{MC}(m,c)\tP_{\bX|MC}(\bx|m,c)\;p_{\bY\bZ_S|\bX}(\by,\bz|\bx)\tP_{\bXh|\bY C}(\bxh|\by,c).
\end{align}

{\it{Protocol B (Main problem assisted with common randomness):}} This protocol is defined as the channel model in Figure \ref{fig:sysmodel}, with an addition of a common randomness $C$ that is uniform over $\llbracket 1,2^{n\tilde{R}_s}\rrbracket$, independent from all other variables, and known at all terminals. In fact, the assumed common randomness represents the random nature in generating the codebook, which is known at all nodes. At the end of the proof, we eliminate the assumed common randomness from the channel model in this protocol by conditioning on a certain instance of it. The encoder and decoder in this protocol are defined as in (\ref{eq:P_tilde_2}), i.e.,  $P_{\bX|MC}=\tP_{\bX|MC}$ and $P_{\bXh|\bY C}=\tP_{\bXh|\bY C}$. The induced joint distribution for this protocol is given by
\begin{align}
\label{eq:P_1}
P_{MC\bX\bY\bZ_S\bXh}(m,c,\bx,\by,\bz,\bxh)=p_M^Up_C^U \tP_{\bX|MC}(\bx|m,c)\;p_{\bY\bZ_S|\bX}(\by,\bz|\bx)\tP_{\bXh|\bY C}(\bxh|\by,c).
\end{align}

The induced joint distributions in (\ref{eq:P_tilde_2}) and (\ref{eq:P_1}) are random due to the random binning of $\bX^n$. Note that we have ignored the random variables $\hat{M}$ from the induced joint distributions at this stage. We will introduce them later to the joint distributions as deterministic functions of the random vectors $\bXh^n$, after fixing the binning functions. 

The remaining steps of the proof are outlined as follows: 
\begin{enumerate}[(i)]
\item We derive a condition on the rates $R_s$ and $\tilde{R}_s$ such that the two induced joint distributions (\ref{eq:P_tilde_2}) and (\ref{eq:P_1}) are close in the total variation distance sense, when averaged over the random binning.
 
\item We then use Slepian-Wolf source coding theorem \cite{slepian1973noiseless,cover1975proof} to derive a condition on the rate $\tilde{R}_s$ such that the decoding of $\bXh$ in protocol A is reliable. 

\item Next, for protocol A, we derive another condition on the rates $R_s$ and $\tilde{R}_s$ such that the probability, with respect to the random binning, that for any $S\in\setS$, the messages $M$ and $C$ are uniform and independent from the wiretapper's observation $\bZ_S^n$, goes to one as $n$ goes to infinity, i.e., protocol A is secure.  
  
\item We use the closeness of the two induced distributions for the two protocols to show that the aforementioned reliability and secrecy properties hold for protocol B as well. 

\item Finally, we eliminate the common randomness $C$ from the channel model in protocol B by showing that the reliability and secrecy constraints still hold when we condition on a certain instance of $C$, i.e., $C=c^{*}$. 
\end{enumerate}

Note that, for the secrecy constraint, we have required the independence of the assumed common random $C$ from both $M$ and $\bZ_S^n$ so that when we condition over an instance of $C$, the independence of $M$ and $\bZ_S^n$ is not affected. Before continuing with the proof, we state the following lemmas.

\subsection{Useful Lemmas}
Lemma \ref{lemma1} is a {\it{one-shot}} result, which provides an exponential decay rate for the average, over the random binning, of the total variation distance between the two induced  distributions from the two protocols. We utilize this lemma to show a convergence in probability result that allows converting the secrecy property from protocol A to protocol B. A result similar to Lemma \ref{lemma1} was derived in \cite[Appendix A]{yassaee2014achievability} which does not provide the required convergence rate, hence the need for Lemma \ref{lemma1}.
\begin{lemma}\label{lemma1}
Let the source $X\triangleq\{\setX,p_X\}$ be randomly binned into $M=\setB_1(X)$ and $C=\setB_2(X)$, where $\setB_1$ and $\setB_2$ are uniform over $\llbracket 1,\tM\rrbracket$ and $\llbracket 1,\tC\rrbracket$, respectively. Let $\setB\triangleq \{\setB_1(x),\setB_2(x)\}_{x\in\setX}$, and for $\gamma>0$, define 
\begin{align}
\label{eq:D_gamma}
\setD_{\gamma}\triangleq \left\{x\in \setX: \log\frac{1}{p_X(x)}>\gamma\right\}.
\end{align}
Then, we have 
\begin{align}
\label{eq:lemma1}
\E_{\setB}\left(\V\left(P_{MC},p_M^U p_C^U\right)\right)\leq \Prob\left(X\notin \setD_{\gamma}\right)+\frac{1}{2}\sqrt{\tM\tC2^{-\gamma}},
\end{align}
where $P$ is the induced distribution over $M$ and $C$.
\end{lemma}
\begin{Proof}
The proof is provided in Appendix \ref{appendix_A}. 
\end{Proof}

Lemma \ref{lemma2} below is again a one-shot result which  provides a {\it{doubly-exponential}} decay rate for the probability of failure of achieving the secrecy property for protocol A, for a fixed choice of the subset $S$.  This lemma is needed, along with the union bound, to guarantee secrecy against the exponentially many possibilities of the tapped subset $S$.  
\begin{lemma}\label{lemma2}
Let $X\triangleq\{\setX,p_X\}$ and $\{Z_{S}\}\triangleq\left\{\setZ,p_{Z_S}, S\in\setS \right\}
$ be two correlated sources with $|\setX|,|\setZ|$, and $|\setS|<\infty$, where $\{Z_S\}_{S\in\setS}$ is a compound source whose distribution is known to belong to the finite class $\{p_{Z_S}\}_{S\in\setS}$. Let $X$ be randomly binned into the bin indices $M$ and $C$ as in Lemma \ref{lemma1}. For $\gamma>0$ and any $S\in\setS$, define
\begin{align}
\label{eq:D_gamma_S}
\setD_{\gamma}^S\triangleq \left\{(x,z)\in \setX\times \setZ: \log \frac{1}{p_{X|Z_S}(x|z)}>\gamma\right\}.
\end{align}
If there exists $\delta\in\left(0,\frac{1}{2}\right)$ such that for all $S\in\setS$, $\Prob_{p_{XZ_S}}\left((X,Z_S)\in\setD_\gamma^S\right)\geq 1-\delta^2$, then, we have, for every $\epsilon_1\in[0,1]$, that
\begin{align}\label{eq:lemma2}
\Prob_{\setB}\left(\underset{S\in\setS}\max\;\D\left(P_{MCZ_S}||p_M^Up_C^Up_{Z_S}\right)\geq\tilde{\epsilon}\right)\leq |\setS||\setZ|\exp\left(\frac{-\epsilon_1^2 (1-\delta)2^{\gamma}}{3\tM\tC}\right), 
\end{align}
where $\tilde{\epsilon}=\epsilon_1+(\delta+\delta^2)\log(\tM\tC)+H_b(\delta^2)$, $H_b$ is the binary entropy function, and $P$ is the induced distribution over $M,C,$ and $Z_S$.
\end{lemma}
\begin{Proof}
The proof of Lemma \ref{lemma2} is given in Appendix \ref{appendix_B}. 
\end{Proof}

The selection lemma below is used to show the existence of a binning realization  such that both the secrecy and reliability properties hold for protocol B. It is also used to eliminate the common randomness $C$ from the channel model in protocol B. 
\begin{lemma}(Selection Lemma)\cite[Lemma 2.2]{bloch2011physical}:\label{lemma3}\\
Let $A_1,A_2,\cdots,A_n$ be a sequence of random variables where $A_n\in\setA^n$, and let $\setF_n=\left\{f_{1,n},\cdots,f_{M,n}\right\}$ be a finite set of bounded functions $f_{i,n}:\setA^n\mapsto \mathbb{R}^{+},\;i\in\llbracket 1,M \rrbracket$, such that $|\setF_n|=M$ does not depend on $n$, and 
\begin{align}
\limitn\E_{\setA^n}\left(f_{i,n}(A_n)\right)=0\;\;\text{ for all }i\in\llbracket 1,M\rrbracket.
\end{align}
Then, there exists a specific realization $\{a^*_n\}$ of the sequence $\{A_n\}$ such that 
\begin{align}
\limitn f_{i,n}(a^*_n)=0\;\;\text{ for all }i\in\llbracket 1,M\rrbracket.
\end{align} 
\end{lemma} 
 
The following Lemma states two properties of the total variation distance, which we utilize through the achievability proof. 
\begin{lemma}(Properties of Total Variation Distance)\cite[Lemmas V.1 and V.2]{cuff2013distributed}:\label{lemma4}\\
Consider the joint distributions $p_{X,Y}=p_{X}p_{Y|X}$ and $q_{X,Y}=q_Xq_{Y|X}$, defined on the same probability space. Then, we have,
\begin{align}
\label{eq:V_prop_1}
&\V(p_X,q_X)\leq \V(p_{X,Y},q_{X,Y})\\
\label{eq:V_prop_2}
&\V(p_Xp_{Y|X},q_Xp_{Y|X})=\V(p_X,q_X).
\end{align}
\end{lemma}

In order to apply Lemmas \ref{lemma1} and \ref{lemma2} to protocol A, we use Hoeffding's inequality, which is stated in the following Lemma.
\begin{lemma}(Hoeffding's Inequality) \cite[Theorem 2]{hoeffding1963probability}:\label{lemma5}\\
Let $X_1,X_2,\cdots,X_n$ be independent random variables with $X_i\in[0,b]$ for all $i\in\llbracket 1,n\rrbracket$, and let $\bar{m}=\frac{1}{n}\sum_{i=1}^n\E(X_i)$. Then, for $\epsilon>0$, we have
\begin{align}
\label{eq:hoeffding}
\Prob\left(\frac{1}{n}\sum_{i=1}^n X_i\leq (1-\epsilon)\bar{m}\right)\leq \exp\left(\frac{-2\epsilon^2\bar{m}^2}{b^2}n\right).
\end{align}
\end{lemma}

\subsection{Proof}
First, we apply Lemma \ref{lemma1} to protocol A. In Lemma \ref{lemma1}, set $X=\bX^n$, $\tM=2^{nR_s}$, $\tC=2^{n\tilde{R}_s}$, $\setB=\setB_n=\left\{\setB_{1,n}(\bx),\setB_{2,n}(\bx)\right\}_{\bx\in\setX^n}$, and $\gamma=n(1-\epsilon_2)H(X)$, where $\epsilon_2>0$ and $\bX^n$ is defined as in protocol A, i.e., is an i.i.d. sequence. Without loss of generality, we assume that for all $x\in\setX$, we have $p_X(x)>0$. Let $p_{\rm{min}}=\min_{x\in\setX}\;p_X(x)$, where the minimum exists since the input alphabet $\setX$ is finite\footnote{If the input alphabet $\mathcal{X}$ is infinite, $\min_{x\in\setX}\;p_X(x)$ might not exist. As a result, there might not be a finite upper bound on the random variables $\log\frac{1}{p_X(X_i)}$. In such a case, Hoeffding inequality can not be applied.}. Thus, the random variables $\log\frac{1}{p_X(X_i)}, i\in\llbracket 1,n \rrbracket,$ are i.i.d. and each is bounded by the interval $\left[0,b_{\rm{max}}\right]$, where $b_{\rm{max}}=\log \frac{1}{p_{{\rm{min}}}}$. 

We also have that $\bar{m}=\frac{1}{n}\sum_{i=1}^n \E_{p_X}\left(\log\frac{1}{p_X(X_i)}\right)=H(X)$. Using Hoeffding's inequality in (\ref{eq:hoeffding}), we have, for any $\epsilon_2>0$, that
\begin{align}
\label{eq:achievability_1}
\Prob&\left(\bX\notin\setD_{\gamma}\right)=\Prob_{p_{\bX}}\left(\log \frac{1}{p_{\bX}(\bX)}\leq \gamma \right)\\
\label{eq:achievability_2}
&=\Prob_{p_X}\left(\frac{1}{n}\sum_{i=1}^n \log \frac{1}{p_{X}(X_i)}\leq (1-\epsilon_2)H(X)\right)\\
\label{eq:achievability_3}
&\leq \exp\left(\frac{-2\epsilon_2^2 H(X)^2}{b_{\rm{max}}^2}n\right)= \exp(-\beta_1 n),
\end{align} 
where $\beta_1=\frac{2\epsilon_2^2 H(X)^2}{b_{\rm{max}}^2}>0$. 

By substituting the choices for $\tM,\tC,\gamma$ and (\ref{eq:achievability_3}) in (\ref{eq:lemma1}), we have, as long as $R_s+\tilde{R}_s<(1-\epsilon_2)H(X)$, that 
\begin{align}
\label{eq:close_1}
\E_{\setB_n}\big(\V(\tP_{MC},p_M^Up_C^U)\big)\leq 2\exp(-\beta n),
\end{align}
where $\beta_2=\frac{\ln 2}{2}\left((1-\epsilon_2)H(X)-R_s-\tilde{R}_S\right)$ and $\beta=\min\{\beta_1,\beta_2\}>0$. By applying (\ref{eq:V_prop_2}) to (\ref{eq:P_tilde_2}) and (\ref{eq:P_1}), and using (\ref{eq:close_1}), we have
\begin{align}
\label{eq:close_2}
\E_{\setB_n}\left(\V\left(\tP_{MC\bX\bY\bZ_S\bXh},P_{MC\bX\bY\bZ_S\bXh}\right)\right)=\E_{\setB_n}\left(\V\left(\tP_{MC},p_M^U p_C^U\right)\right)\leq 2\exp(-\beta n).
\end{align}

Consider Slepian-Wolf decoder for protocol A. As long as $\tilde{R}_s\geq H(X|Y)$, we have \cite[Theorem 10.1]{el2011network}
\begin{align}
\label{eq:achievability_4}
\lim_{n\rightarrow\infty}\E_{\setB_n}\left(\Prob_{\tP}(\bXh\neq \bX)\right)=0.
\end{align} 

Next, we observe
\begin{align}
\label{eq:achievability_5}
\nonumber \E_{\setB_n}&\left(\V\left(\tP_{MC\bX\bY\bZ_S\bXh},\tP_{MC\bX\bY\bZ_S}\mathbbm{1}\{\bXh=\bX\}\right)\right)\\
&=\E_{\setB_n}\underset{\begin{subarray}{c} m,c,\bx,\by,\bz,\bxh:\\\tP(m,c,\bx,\by,\bz,\bxh)>\tP(m,c,\bx,\by,\bz)\mathbbm{1}\{\bxh=\bx\}\end{subarray}}\sum \left[\tP(m,c,\bx,\by,\bz,\bx)-\tP(m,c,\bx,\by,\bz)\mathbbm{1}\{\bxh=\bx\}\right]\\
\label{eq:achievability_6}
&=\E_{\setB_n}\underset{m,c,\bx,\by,\bz,\bxh:\;\bxh \neq \bx}\sum\tP(m,c,\bx,\by,\bz,\bx)=\E_{\setB_n}\left(\Prob_{\tP}(\bXh\neq \bX)\right).
\end{align}
Equation (\ref{eq:achievability_6}) follows because $\tP(m,c,\bx,\by,\bz,\bxh)>\tP(m,c,\bx,\by,\bz)\mathbbm{1}\{\bxh=\bx\}$ holds if and only if $\mathbbm{1}\{\bxh=\bx\}=0$, where $\tP(m,c,\bx,\by,\bz,\bxh)$ factorizes as $\tP(m,c,\bx,\by,\bz,\bxh)=\tP(m,c,\bx,\by,\bz)\tP(\bxh|\by,c)$ and $\tP(\bxh|\by,c)\leq 1$. Thus, using (\ref{eq:achievability_4}) and (\ref{eq:achievability_6}), we have that 
\begin{align}
\label{eq:reliability_A}
\limitn\E_{\setB_n}\left(\V\left(\tP_{MC\bX\bY\bZ_S\bXh},\tP_{MC\bX\bY\bZ_S}\mathbbm{1}\{\bXh=\bX\}\right)\right)=0,
\end{align}
as long as $\tilde{R}_s\geq H(X|Y)$.

Now, we apply Lemma \ref{lemma2} to protocol A. In Lemma \ref{lemma2}, set $X=\bX^n$, $\tM=2^{nR_s}$, $\tC=2^{n\tilde{R}_s}$, $\setB=\setB_n$, $Z_S=\bZ_S^n$, for all $S\in\setS$, and $\gamma=n(1-\tilde{\epsilon}_2)(1-\alpha)H(X|V)$, where $\tilde{\epsilon}_2>0$ and $\bX^n,\bZ_S^n,\setS$ are defined as in protocol A. In order to calculate $\Prob_{p_{\bX\bZ_S}}\left((\bX,\bZ_S)\notin\setD_\gamma^S\right)$, we only need to consider the pairs $(\bx,\bz)$ such that $p_{\bX|\bZ_S}(\bx|\bz)>0$, since all the pairs $(\bx,\bz)$ with $p_{\bX|\bZ_S}(\bx|\bz)=0$ belong to $\setD_\gamma^S$, by the definition of $\setD_\gamma^S$ in (\ref{eq:D_gamma_S}). Since the sequence $\bX$ is i.i.d. and the channel $p_{V|X}$ is memoryless, we have, for all $(\bx,\bz)$ with $p_{\bX|\bZ_S}(\bx|\bz)>0$, that  
\begin{align}
\label{eq:achievability_7}
p_{\bX|\bZ_S}(\bx|\bz)&=p_{\bX_S\bX_{S^c}|\bX_S\bV_{S^c}}(\bx_S,\bx_{S^c}|\bx_S,\bv_{S^c})\\
\label{eq:achievability_8}
&=p_{\bX_{S^c}|\bV_{S^c}}(\bx_{S^c}|\bv_{S^c})=\prod_{i\in S^c}p_{X|V}(x_i|v_i).
\end{align}

Once again, using Hoeffding's inequality, we have, for all $S\in\setS$,
\begin{align}
\label{eq:achievability_9}
\Prob_{p_{\bX\bZ_S}}&\left((\bX,\bZ_S)\notin\setD_\gamma^S\right)=\Prob_{p_{\bX\bZ_S}}\left(p_{\bX|\bZ_S}(\bX|\bZ_S)>0, \;\log \frac{1}{p_{\bX|\bZ_S}(\bX|\bZ_S)}\leq \gamma \right)\\
\label{eq:achievability_10}
&=\Prob_{p_{X|V}}\left(\frac{1}{n-\mu}\sum_{i\in S^c}\log\frac{1}{p_{X|V}(X_i|V_i)}\leq (1-\tilde{\epsilon}_2)H(X|V)\right)\\
\label{eq:achievability_11}
&\leq \exp\left(-\beta_2 (1-\alpha)n\right)=\delta^2,
\end{align}
where $\beta_2>0$, and (\ref{eq:achievability_10}) follows from (\ref{eq:achievability_8}). From (\ref{eq:achievability_11}), $\limitn\delta^2=0$, and hence, for sufficiently large $n$ , we have $\delta^2\in\left(0,\frac{1}{4}\right)$. Thus, the conditions in Lemma \ref{lemma2} are satisfied.

Note that $\limitn n(\delta+\delta^2)=0$, and $\limitn H_b(\delta^2)=H_b(\limitn \delta^2)=0$ since $H_b$ is a continuous function. Thus, 
\begin{align}
\label{eq:achievability_11_1}
\limitn\tilde{\epsilon}=\epsilon_1+(R_s+\tilde{R}_s)\limitn n(\delta+\delta^2)+\limitn H_b(\delta^2)=\epsilon_1.
\end{align}
By substituting the choices for $\tM,\tC,\gamma$, and $|\setS||\setZ^n|\leq \exp\left(n\left[\ln2+\ln\left(|\setX|+|\setV|\right)\right]\right)$ in (\ref{eq:lemma2}), and using (\ref{eq:achievability_11_1}), we have that, for all $\epsilon_1,\epsilon'_1>0$ and $\tilde{\epsilon}=\epsilon_1+\epsilon_1'$, there exist $n^*\in\mathbb{N}$ and $\phi(\epsilon_1),\kappa>0$ such that, for all $n\geq n^*$,
\begin{align}
\label{eq:secrecy_A_3}
\Prob_{\setB_n}\left(\underset{S\in\setS}\max\;\D\left(\tP_{MC\bZ_S}||p_M^Up_C^Up_{\bZ_S}\right)\geq \tilde{\epsilon}\right)\leq \exp\left(-\phi(\epsilon_1)e^{\kappa n}\right),
\end{align}
as long as $R_s+\tilde{R}_s<(1-\tilde{\epsilon}_2)(1-\alpha) H(X|V)$.

Take $r>0$ and let $D_n=\max_S\D(\tP_{MC\bZ_S}||p_M^Up_C^Up_{\bZ_S})$ and $\setK_n\triangleq\left\{D_n\geq r\right\}$. Using (\ref{eq:secrecy_A_3}), we have that $\sum_{n=1}^\infty \Prob_{\setB_n}(\setK_n)<\infty$. Thus, using the first Borel-Cantelli lemma yields
\begin{align}
\Prob_{\setB_n}\left(\setK_n \text{ infinitely often (i.o.)}\right)=0. 
\end{align}
This implies that, for all $r>0$, $\Prob_{\setB_n}\left(\{D_n<r\} \text{ i.o.}\right)=1$, i.e., the sequence $D_n$ converges to zero almost surely. Thus, the sequence $D_n$ converges to zero in probability as well. We conclude that, for $R_s+\tilde{R}_s<(1-\tilde{\epsilon}_2)(1-\alpha)H(X|V)$, we have
\begin{align}
\label{eq:secrecy_A_4}
\limitn\Prob_{\setB_n}\left(\underset{S\in\setS}\max\;\D\left(\tP_{MC\bZ_S}||p_M^Up_C^Up_{\bZ_S}\right)>0\right)=0.
\end{align}
That is, protocol A is secure.

Next, we deduce that protocol B is also reliable and secure when $\tilde{R}_s\geq H(X|Y)$ and $R_{s}+\tilde{R}_s<(1-\tilde{\epsilon}_2)(1-\alpha)H(X|V)$. First, we show that the reliability in (\ref{eq:reliability_A}) holds for protocol B as well. We have 
\begin{align} 
\label{eq:achievability_12}
\nonumber \V&\left(P_{MC\bX\bY\bZ_S\bXh},P_{MC\bX\bY\bZ_S}\mathbbm{1}\{\bXh=\bX\}\right)\\
&\leq\V\left(P_{MC\bX\bY\bZ_S\bXh},\tP_{MC\bX\bY\bZ_S\bXh}\right)+\V\left(\tP_{MC\bX\bY\bZ_S\bXh}, P_{MC\bX\bY\bZ_S}\mathbbm{1}\{\bXh=\bX\}\right)\\
\nonumber\\
\label{eq:achievability_13}
&\nonumber \leq\V\left(P_{MC\bX\bY\bZ_S\bXh},\tP_{MC\bX\bY\bZ_S\bXh}\right)+\V\left(\tP_{MC\bX\bY\bZ_S\bXh}, \tP_{MC\bX\bY\bZ_S}\mathbbm{1}\{\bXh=\bX\}\right)\\
&\qquad\qquad\qquad\qquad +\V\left(\tP_{MC\bX\bY\bZ_S}\mathbbm{1}\{\bXh=\bX\},P_{MC\bX\bY\bZ_S}\mathbbm{1}\{\bXh=\bX\}\right)\\
\label{eq:achievability_14}
&=\V\left(\tP_{MC\bX\bY\bZ_S\bXh}, \tP_{MC\bX\bY\bZ_S}\mathbbm{1}\{\bXh=\bX\}\right)+2\V\left(\tP_{MC},p_M^U p_C^U\right),
\end{align}
where (\ref{eq:achievability_12}) and (\ref{eq:achievability_13}) follow from the triangle inequality, and (\ref{eq:achievability_14}) follows since (\ref{eq:P_tilde_2}), (\ref{eq:P_1}) and (\ref{eq:V_prop_2}) imply that 
\begin{align}
\label{eq:achievability_15}
\nonumber\V\left(P_{MC\bX\bY\bZ_S\bXh},\tP_{MC\bX\bY\bZ_S\bXh}\right)&=\V\left(\tP_{MC\bX\bY\bZ_S}\mathbbm{1}\{\bXh=\bX\},P_{MC\bX\bY\bZ_S}\mathbbm{1}\{\bXh=\bX\}\right)\\
&=\V\left(\tP_{MC},p_M^U p_C^U\right).
\end{align}
Substituting (\ref{eq:close_1}) and (\ref{eq:reliability_A}) in (\ref{eq:achievability_14}) yields 
\begin{align}
\label{eq:reliability_B_1}
\limitn\E_{\setB_n}\left(\V\left(P_{MC\bX\bY\bZ_S\bXh},P_{MC\bX\bY\bZ_S}\mathbbm{1}\{\bXh=\bX\}\right)\right)=0.
\end{align} 

Second, we show that the secrecy property in (\ref{eq:secrecy_A_4}) holds for protocol B. Using the union bound, we have 
\begin{align}
\label{eq:achievability_16}
\nonumber \Prob_{\setB_n}&\left(\max_{S\in\setS}\D\left(P_{MC\bZ_S}||p_M^Up_C^Up_{\bZ_S}\right)>0\right)\\
\nonumber &\leq \Prob_{\setB_n}\left(\max_{S\in\setS} \D\left(P_{MC\bZ_S}||p_M^Up_C^Up_{\bZ_S}\right)>0,\;\V(\tP_{MC},p_M^Up_C^U)>0\right)\\
&\qquad +\Prob_{\setB_n}\left(\max_{S\in\setS} \D\left(P_{MC\bZ_S}||p_M^Up_C^Up_{\bZ_S}\right)>0,\;\V(\tP_{MC},p_M^Up_C^U)=0\right)\\
\label{eq:achievability_17}
&\leq \Prob_{\setB_n}\left(\V(\tP_{MC},p_M^Up_C^U)>0\right)+\Prob_{\setB_n}\left(\max_{S\in\setS} \D\left(\tP_{MC\bZ_S}||p_M^Up_C^Up_{\bZ_S}\right)>0\right).
\end{align}
Equation (\ref{eq:achievability_17}) follows since $\V(\tP_{MC},p_M^U p_C^U)=0$ if and only if $\tP_{MC}(m,c)=p_M^U p_C^U$ for all $m$ and $c$, and hence $P_{MC\bZ_S}=p_M^Up_C^UP_{\bZ_S|MC}=\tP_{MC}\tP_{\bZ_S|MC}=\tP_{MC\bZ_S}$, where
\begin{align} 
\label{eq:achievability_18}
P_{\bZ_S|MC}(\bz|m,c)=\sum_{\bx\in\setX^n} p_{\bZ_S|\bX}(\bz|\bx)\tP_{\bX|MC}(\bx|m,c)=\tP_{\bZ_S|MC}(\bz|m,c).
\end{align}

Using the exponential decay in (\ref{eq:close_1}) and Markov inequality, we have, for any $r>0$, that
\begin{align}
\label{eq:achievability_19}
\sum_{n=1}^\infty \Prob_{\setB_n}\left(\V(\tP_{MC},p_M^U p_C^U)>r\right)&\leq \frac{1}{r} \sum_{n=1}^\infty\E_{\setB_n}\left(\V(\tP_{MC},p_M^U p_C^U)\right)\\
\label{eq:achievability_20}
&\leq \frac{2}{r}\sum_{n=1}^\infty  \exp(-\beta n)<\infty, 
\end{align}
where $\beta>0$. Thus, using the Borel-Cantelli lemma, as in the derivation for (\ref{eq:secrecy_A_4}), we have
\begin{align}
\label{eq:achievability_21}
\limitn\Prob_{\setB_n}\left(\V(\tP_{MC},p_M^Up_C^U)>0\right)=0.
\end{align} 
By substituting (\ref{eq:secrecy_A_4}) and (\ref{eq:achievability_21}) in (\ref{eq:achievability_17}), we get
\begin{align}
\label{eq:secrecy_B_1}
\limitn\Prob_{\setB_n}&\left(\max_{S\in\setS} \D\left(P_{MC\bZ_S}||p_M^Up_C^Up_{\bZ_S}\right)>0\right)=0.
\end{align}

Now, we show the existence of a binning realization, and hence an encoder and decoder, such that the reliability and secrecy properties, in (\ref{eq:reliability_B_1}) and (\ref{eq:secrecy_B_1}), hold for protocol B. By applying Lemma \ref{lemma3} to the random sequence $\{\setB_n\}_{n\geq 1}$ and the functions $\V\left(P_{MC\bX\bY\bZ_S\bXh},P_{MC\bX\bY\bZ_S}\mathbbm{1}\{\bXh=\bX\}\right)$, $\mathbbm{1}\left\{\max_{S\in\setS} \D(P_{MC\bZ_S}||p_M^Up_C^Up_{\bZ_S})>0\right\}$, while using (\ref{eq:reliability_B_1}) and (\ref{eq:secrecy_B_1}), there exists a sequence of binning realizations $\bold{b}_n^*=(b_{1,n}^*,b_{2,n}^*)$, with a corresponding joint distribution $p^*$ for protocol B, such that
\begin{align}
\label{eq:reliability_B_2}
&\limitn\V\left(p^*_{MC\bX\bY\bZ_S\bXh},p^*_{MC\bX\bY\bZ_S}\mathbbm{1}\{\bXh=\bX\}\right)=0,\\
\label{eq:secrecy_B_2}
&\limitn\mathbbm{1}\left\{\max_{S\in\setS} \D(p^*_{MC\bZ_S}||p_M^Up_C^Up_{\bZ_S})>0\right\}=0,
\end{align}
where $M=b^{*}_{1,n}(\bX^n)$ and $C=b^{*}_{2,n}(\bX^n)$. 

Next, we introduce the random variable $\hat{M}$ to the two joint distributions in (\ref{eq:reliability_B_2}), where $\hat{M}$ is a deterministic function of the random sequence $\bXh^n$, i.e., $p^*_{\hat{M}|\bXh}(\hat{m}|\bxh)=\mathbbm{1}\left\{\hat{m}=b_{1,n}^*(\bxh)\right\}$. Using (\ref{eq:V_prop_2}) and (\ref{eq:reliability_B_2}), we have
\begin{align}
\label{eq:achievability_22}
\nonumber \limitn\V&\left(p^*_{MC\bX\bY\bZ_S\bXh\hat{M}},p^*_{MC\bX\bY\bZ_S}\mathbbm{1}\{\hat{M}=M\}\right)\\
&=\limitn\V\left(p^*_{MC\bX\bY\bZ_S\bXh}\mathbbm{1}\big\{\hat{M}=b_{1,n}^*(\bXh)\big\},p^*_{MC\bX\bY\bZ_S}\mathbbm{1}\{\bXh=\bX\}\mathbbm{1}\big\{\hat{M}=b_{1,n}^*(\bXh)\big\}\right)\\
\label{eq:achievability_23}
&=\limitn\V\left(p^*_{MC\bX\bY\bZ_S\bXh},p^*_{MC\bX\bY\bZ_S}\mathbbm{1}\{\bXh=\bX\}\right)=0,
\end{align}
where (\ref{eq:achievability_22}) follows since $p^*_{\hat{M}|MC\bX\bY\bZ_S\bXh}=p^*_{\hat{M}|\bXh}=\mathbbm{1}\big\{\hat{M}=b_{1,n}^*(\bXh)\big\}$, and that $\hat{M}=M$ if and only if $\bXh=\bX$ and $\hat{M}=b_{1,n}^*(\bXh)$. We then have
\begin{align}
\label{eq:achievability_24}
\limitn\E_C&\left(\Prob_{p^*}(\hat{M}\neq M|C)\right)
=\limitn\sum_{c} p_C^U \sum_{m,\hat{m}:\;\hat{m}\neq m} p^*_{M\hat{M}|C}(m,\hat{m}|c)\\
\label{eq:achievability_25}
&=\limitn\sum_{m,\hat{m},c:\;\hat{m}\neq m}p^*_{M\hat{M}C}(m,\hat{m},c)\\
\label{eq:achievability_26}
&=\limitn\sum_{m,\hat{m},c:\;\;p^*(m,\hat{m},c)>p_M^U p_C^U \mathbbm{1}\{\hat{m}=m\}}\left[p^*_{M\hat{M}C}(m,\hat{m},c)-p_M^Up_C^U \mathbbm{1}\{\hat{m}=m\}\right]\\
\label{eq:achievability_27}
&=\limitn\V\left(p^*_{M\hat{M}C},p_M^U p_C^U\mathbbm{1}\{\hat{M}=M\}\right)\\
\label{eq:reliability_B_3}
&=\limitn\V\left(p^*_{MC\bX\bY\bZ_S\bXh\hat{M}},p^*_{MC\bX\bY\bZ_S}\mathbbm{1}\{\hat{M}=M\}\right)=0.
\end{align}
Equation (\ref{eq:achievability_26}) follows because $p^*_{M\hat{M}C}>p_M^U p_C^U \mathbbm{1}\{\hat{M}=M\}$ if and only if $\mathbbm{1}\{\hat{M}=M\}=0$ where $p^*_{M\hat{M}C}$ factorizes as $p_M^U p_C^U p^*_{\hat{M}|MC}$ and $p^*_{\hat{M}|MC}\leq 1$, while equation (\ref{eq:reliability_B_3}) follows from (\ref{eq:V_prop_2}) and (\ref{eq:achievability_23}). 

We also have that
\begin{align}
\nonumber\Prob_{C}&\left(\max_S \D(p^*_{M\bZ_S|C}||p_M^Up^*_{\bZ_S|C})>0\right)\\
\label{eq:achievability_29}
\nonumber &\leq\Prob_C\left(\max_S \D\left(p^*_{M\bZ_S|C}||p_M^Up^*_{\bZ_S|C}\right)>0, \text{ and }\max_S \D\left(p^*_{MC\bZ_S^n}||p_M^Up_C^Up_{\bZ_S}\right)=0\right)\\
&\qquad + \Prob_C\left(\max_S \D\left(p^*_{M\bZ_S|C}||p_M^Up^*_{\bZ_S|C}\right)>0, \text{ and }\max_S \D\left(p^*_{MC\bZ_S^n}||p_M^Up_C^Up_{\bZ_S}\right)>0\right)\\
\label{eq:achievability_30}
\nonumber &\leq \Prob_C\left(\max_S \D\left(p^*_{M\bZ_S|C}||p_M^Up^*_{\bZ_S|C}\right)>0, \text{ and } \forall S,\;p^*_{MC\bZ_S}(m,c,\bz)=p_M^Up_C^Up_{\bZ_S}(\bz), \forall m,c,\bz\right)\\
&\qquad \qquad +\Prob_C\left(\max_S \D(p^*_{MC\bZ_S^n}||p_M^Up_C^Up_{\bZ_S})>0\right)\\
\label{eq:achievability_31}
&=\mathbbm{1}\left\{\max_S \D(p^*_{MC\bZ_S^n}||p_M^Up_C^Up_{\bZ_S})>0\right\},
\end{align}
where (\ref{eq:achievability_30}) follows since $\max_S \D\Big(p^*_{MC\bZ_S^n}||p_M^Up_C^Up_{\bZ_S}\Big)=0$, if and only if, for all $S\in\setS$, $p^*_{MC\bZ_S}(m,c,\bz)=p_M^Up_C^Up_{\bZ_S}(\bz)$ for all $m,c,$ and $\bz$. (\ref{eq:achievability_31}) follows because the first probability term on the right hand side of (\ref{eq:achievability_30}) is equal to zero. Thus, using (\ref{eq:secrecy_B_2}), we get
\begin{align}
\label{eq:secrecy_B_3}
\limitn\Prob_{C}\left(\max_{S\in\setS} \D(p^*_{M\bZ_S|C}||p_M^Up^*_{\bZ_S|C})>0\right)=0.
\end{align}

Let us express the random variable $C$ as an explicit function of $n$, i.e., $C=C_n=b_{2,n}^*(\bX^n)$. In order to eliminate $C_n$ from the channel model in protocol B, we apply Lemma \ref{lemma3} to the random sequence $\{C_n\}_{n\geq 1}$ and the functions $\Prob_{p^*}\left(\hat{M}\neq M|C_n\right)$, $\mathbbm{1}\left\{\max_{S\in\setS} \D\left(p^*_{M\bZ_S|C_n}||p_M^Up^*_{\bZ_S|C_n}\right)>0\right\}$, while using (\ref{eq:reliability_B_3}) and (\ref{eq:secrecy_B_3}), which implies that there exists at least one realization $\{c_n^{*}\}$ such that 
\begin{align}
\label{eq:reliability_B_4}
&\limitn\Prob_{p^*}\left(\hat{M}\neq M|C_n=c_n^*\right)=0,\;\text{ and }\\
\label{eq:secrecy_B_4}
&\limitn \max_{S\in\setS} I_{p^*}\left(M;\bZ_S|C_n=c_n^{*}\right)=0,
\end{align}
where $I_{p^*}$ is the mutual information with respect to the distribution $p^*$. Equation (\ref{eq:secrecy_B_4}) follows because $\limitn\mathbbm{1}\big\{\max_{S\in\setS} \D(p^*_{M\bZ_S|C_n=c_n^*}||p_M^Up^*_{\bZ_S|C_n=c_n^*})>0\big\}=0$ implies that there exists $n'$ large enough such that, for all $n\geq n'$, we have 
\begin{align}
\max_{S} \D\left(p^*_{M\bZ_S|C_n=c_n^*}||p_M^Up^*_{\bZ_S|C_n=c_n^*}\right)=\max_S I_{p^*}\left(M;\bZ_S|C_n=c_n^{*}\right)=0.
\end{align}

Finally, let $\tilde{p}^{*}$ be the induced distribution for protocol A corresponding to $\bold{b}_n^*$. We use $\tilde{p}^{*}_{\bX|M,C_n=c_n^{*}}$ as the encoder and $(\tilde{p}^*_{\bXh|\bY,C_n=c_n^*}, b_{1,n}^*(\bXh))$ as the decoder for the original model. By combining the rate conditions $R_s+\tilde{R}_s<(1-\tilde{\epsilon}_2)(1-\alpha)H(X|V)$, $\tilde{R}_s\geq H(X|Y)$, and taking $\tilde{\epsilon}_2\rightarrow 0$, the rate $R_s=\max_{p_X}[I(X;Y)-I(X;V)-\alpha H(X|V)]$ is achievable. 

So far, we have considered the case $U=X$. Next, we prefix a discrete memoryless channel $p_{X|U}$ to the original channel model in Figure \ref{fig:sysmodel}. The main channel for the new model is $p_{Y|U}$ and the wiretapper channel is described by $p_{X|U}$ and (\ref{eq:Z_S_n}). The proof for this case follows similar steps to the proof above. In particular, for protocol A, we consider the i.i.d. input sequence $\bU^n=[U_1,U_2,\cdots,U_n]$. When we apply Lemma \ref{lemma2} to protocol A, we set $\gamma=n(1-\tilde{\epsilon}_2)[\alpha H(U|X)+(1-\alpha)H(U|V)]$, and for $p_{\bU|\bZ_S}(\bu|\bz)>0$, we have, for any $S\in\setS$, that
\begin{align} 
\label{eq:achievability_32}
p_{\bU|\bZ_S}(\bu|\bz)&=p_{\bU_S\bU_{S^c}|\bX_S\bV_{S^c}}(\bu_S,\bu_{S^c}|\bx_S,\bv_{S^c})\\
\label{eq:achievability_33}
&=p_{\bU_S|\bX_S\bV_{S^c}}(\bu_S|\bx_S,\bv_{S^c})\;p_{\bU_{S^c}|\bU_S \bX_S\bV_{S^c}}(\bu_{S^c}|\bu_S,\bx_S,\bv_{S^c})\\
\label{eq:achievability_34}
&=p_{\bU_S|\bX_S}(\bu_S|\bx_S)\;p_{\bU_{S^c}|\bV_{S^c}}(\bu_{S^c}|\bv_{S^c})\\
\label{eq:achievability_35}
&=\prod_{i\in S}p_{U|X}(u_i|x_i)\prod_{i\in S^c}p_{U|V}(u_i|v_i),
\end{align}
where (\ref{eq:achievability_34}) and (\ref{eq:achievability_35}) follow since the sequences $\bU^n,\bX^n,$ and $\bV^n$ are i.i.d. and the channels $p_{X|U}$ and $p_{V|X}$ are discrete memoryless channels. Using (\ref{eq:achievability_35}), the choice for $\gamma$, and Hoeffding's inequality, the conditions of Lemma \ref{lemma2} are satisfied, and we deduce the rate condition
\begin{align}
\label{eq:achievability_36}
R_s+\tilde{R}_s<(1-\tilde{\epsilon}_2)[\alpha H(U|X)+(1-\alpha)H(U|V)]
\end{align} 
required for secrecy of protocol A. Note that $H(U|X)=H(U|X,V)$ because of the Markov chain $U-X-V$. By combining (\ref{eq:achievability_36}) with the rate condition $\tilde{R}_s\geq H(U|Y)$ required for the Slepian-Wolf decoder, we obtain the achievability of (\ref{eq:Thm1}). The cardinality bound on $\setU$, $|\setU|\leq |\setX|$, follows using \cite[Appendix C]{el2011network}. This completes the achievability proof of Theorem \ref{thm:Thm1}.  
  
\section{Converse}{\label{Converse}}
\begin{figure}
    \centering
	\includegraphics[scale=0.98]{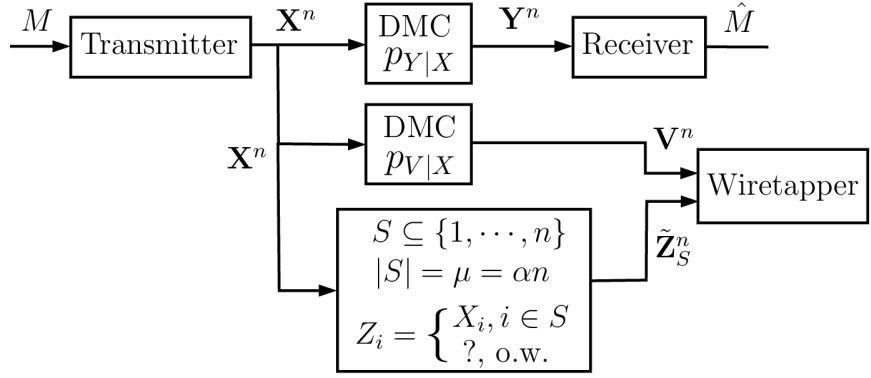}
	\caption{A wiretap channel model whose secrecy capacity is equal to that of Figure \ref{fig:sysmodel}.}
	\label{fig:equiv_model}
\end{figure}

Consider the channel model illustrated in Figure \ref{fig:equiv_model}, where the wiretapper observes the outputs of two independent channels, with $\bX^n$ being the input to both the channels. The first channel to the wiretapper is the DMC $p_{V|X}$ which outputs $\bV^n$. The second channel is the wiretapper channel in the wiretap II channel model, i.e., the wiretapper chooses $S\subseteq \llbracket 1,n\rrbracket$ and observes $\tbZ_S^n=[\tilde{Z}_1^S,\cdots,\tilde{Z}_n^S]$, where $\tilde{Z}_i^S=X_i$ for $i\in S$, and $\tilde{Z}_i^S=?$, i.e., erasures, otherwise. 

We show that, for $0\leq \alpha\leq 1$, the strong secrecy capacity for this channel model, $\mathsf{C}_s^{\rm{EQ}}(\alpha)$, is equal to the strong secrecy capacity of the original channel model, $\mathsf{C}_s(\alpha)$, in (\ref{eq:Thm1}). Since the main channels in the two models are the same, it suffices to show that $I(M;\bZ_S^n)=I(M;\tbZ_S^n\bV^n)$ for all $S\in\setS$, where $\bZ_S^n$ is defined as in (\ref{eq:Z_S_n}). This follows because, for all $S\in\setS$, we have
\begin{align}
\label{eq:converse_1}
H(M|\tbZ_S\bV)&=H(M,\bX|\tbZ_S,\bV)-H(\bX|M,\tbZ_S,\bV)\\
\label{eq:converse_2}
&=H(\bX|\tbZ_S,\bV)+H(M|\bX,\tbZ_S,\bV)-H(\bX|M,\tbZ_S,\bV)\\
\label{eq:converse_3}
&=H(\bX|\tbZ_S,\bV)-H(\bX|M,\tbZ_S,\bV)\\
\label{eq:converse_4}
&=H(\bX_S,\bX_{S^c}|\bX_S,\bV_S,\bV_{S^c})-H(\bX_S,\bX_{S^c}|M,\bX_S,\bV_S,\bV_{S^c})\\
\label{eq:converse_5}
&=H(\bX_{S^c}|\bX_S,\bV_S,\bV_{S^c})-H(\bX_{S^c}|M,\bX_S,\bV_S,\bV_{S^c})\\
\label{eq:converse_6}
&=H(\bX_{S^c}|\bX_S,\bV_{S^c})-H(\bX_{S^c}|M,\bX_S,\bV_{S^c})\\
\label{eq:converse_7}
&=H(\bX|\bZ_S)-H(\bX|M,\bZ_S)\\
\label{eq:converse_8}
&=H(\bX,M|\bZ_S)-H(\bX|M,\bZ_S)=H(M|\bZ_S),
\end{align}
where (\ref{eq:converse_3}) and (\ref{eq:converse_8}) follow because $H(M|\bX)=0$, and (\ref{eq:converse_6}) follows since the channel $p_{V|X}$ is memoryless which results in the Markov chains $\bX_{S^c}-\bX_S\bV_{S^c}-\bV_S$ and $\bX_{S^c}-M\bX_S\bV_{S^c}-\bV_S$.

\begin{figure}
    \centering
	\includegraphics[scale=0.98]{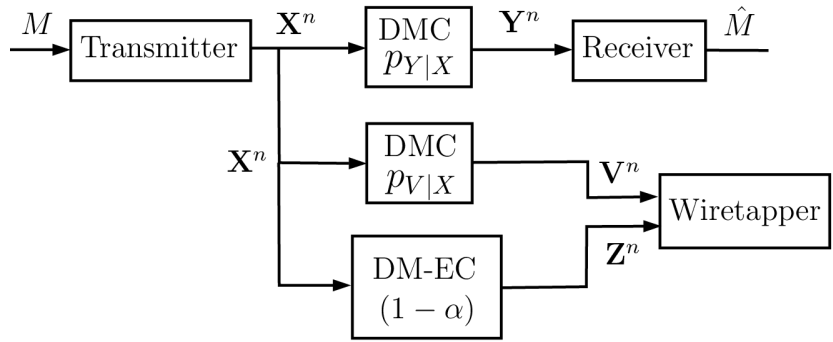}
	\caption{A discrete memoryless equivalent wiretap channel model.}
	\label{fig:DMC_model}
\end{figure}

Next, consider the channel model illustrated in Figure \ref{fig:DMC_model}, which is the same as the channel model in Figure \ref{fig:equiv_model}, except we replace the second channel to the wiretapper with a discrete memoryless erasure channel (DM-EC) with erasure probability $1-\alpha$. The output of the second channel to the wiretapper is $\bZ^n$. For this model, we have the Markov chain $\bV^n-\bX^n-\bZ^n$ since the two channels to the wiretapper are independent. Since the two channels to the wiretapper are discrete memoryless, we have
\begin{align}
\label{eq:converse_9}
\nonumber p_{\bV\bZ|\bX}&(\bv,\bz|\bx)=p_{\bV|\bX}(\bv|\bx)\;p_{\bZ|\bX}(\bz|\bx)\\
&=\prod_{i=1}^n p_{V|X}(v_i|x_i)\;p_{Z|X}(z_i|x_i)=\prod_{i=1}^n p_{VZ|X}(v_i,z_i|x_i).
\end{align}
That is, the combined channel to the wiretapper is a discrete memoryless channel, making the channel model in Figure \ref{fig:DMC_model} a discrete memoryless wiretap channel. The strong secrecy capacity for this model $\mathsf{C}_{s}^{{\rm{EQ2}}}(\alpha)$ is given by 
\begin{align}
\label{eq:converse_10}
\mathsf{C}_{s}^{{\rm{EQ2}}}(\alpha)=\underset{U-X-YVZ}\max[I(U;Y)-I(U;VZ)]^+.
\end{align}

In order to compute $\mathsf{C}_{s}^{{\rm{EQ2}}}(\alpha)$ in (\ref{eq:converse_10}), we define the random variable $\Phi\sim{\rm{Bern}}(\alpha)$ whose $n$ i.i.d. samples represent the erasure process in the DM-EC, where $\Phi=0$ when $Z=X$ and $\Phi=1$ when $Z=?$. Thus, $\Phi$ is determined by $Z$, and hence, the Markov chains $U-X-YVZ$ and $V-X-Z$ imply the Markov chains $U-X-YVZ\Phi$ and $V-X-Z\Phi$. Also, $\Phi$ is independent from $X$, since the erasure process is independent from the input to the channel. Thus, we have
\begin{align}
\label{eq:converse_11}
\nonumber p_{\Phi|UV}(\phi|u,v)&=\sum_{x\in\setX}p_{\Phi X|UV}(\phi,x|u,v)=\sum_{x\in\setX}p_{X|UV}(x|u,v)\;p_{\Phi|XUV}(\phi|x,u,v)\\
&=p_{\Phi}(\phi)\sum_{x\in\setX}p_{X|UV}(x|u,v)=p_{\Phi}(\phi)\\
\label{eq:converse_12}
\nonumber p_{\Phi|V}(\phi|v)&=\sum_{x\in\setX}p_{\Phi X|V}(\phi,x|v)=\sum_{x\in\setX}p_{X|V}(x|v)\;p_{\Phi|XV}(\phi|x,v)\\
&=p_{\Phi}(\phi)\sum_{x\in\setX}p_{X|V}(x|v)=p_{\Phi}(\phi).
\end{align}
where (\ref{eq:converse_11}) and (\ref{eq:converse_12}) follow since $p_{\Phi|XUV}=p_{\Phi|XV}=p_{\Phi|X}=p_{\Phi}$ due to the Markov chains $U-XV-\Phi$ and $V-X-\Phi$, and the independence of $\Phi$ and $X$. Since $p_{\Phi|UV}=p_{\Phi|V}=p_{\Phi}$, then $\Phi$ and $U$ are conditionally independent given $V$. Thus, we have
\begin{align}
\label{eq:converse_13}
I(U;Z|V)&=I(U;Z,\Phi|V)=I(U;Z|\Phi,V)\\
\label{eq:converse_14}
&=\Prob(\Phi=0)I(U;Z|\Phi=0,V)+\Prob(\Phi=1)I(U;Z|\Phi=1,V)\\
\label{eq:converse_15}
&=\alpha I(U;X|V)+(1-\alpha)I(U;?|V)=\alpha I(U;X|V).
\end{align}
Substituting (\ref{eq:converse_15}) in (\ref{eq:converse_10}), we have 
\begin{align}
\label{eq:upper bound}
\mathsf{C}_{s}^{{\rm{EQ2}}}(\alpha)=\underset{U-X-YV}\max[I(U;Y)-I(U;V)-\alpha I(U;X|V)]^+.
\end{align}

Next, we use similar arguments to \cite[Section V-C]{goldfeld2015semantic} to show that $\mathsf{C}_{s}^{{\rm{EQ}}}(\alpha)\leq \mathsf{C}_{s}^{{\rm{EQ2}}}(\alpha)$ for any $0\leq\alpha\leq 1$ and sufficiently large $n$. The idea is that when the number of erasures of the DM-EC in the latter channel model (Figure \ref{fig:DMC_model}) is  greater than or equal to $(1-\alpha)n$, the wiretapper's channel in the former (Figure \ref{fig:equiv_model}) is better than her channel in the latter, since the wiretapper in the former is more capable and encounters a smaller number of erasures. Thus, $\mathsf{C}_{s}^{{\rm{EQ}}}(\alpha)\leq \mathsf{C}_{s}^{{\rm{EQ2}}}(\alpha)$ for this case. The result is established by using Sanov's theorem in method of types \cite[Theorem 11.4.1]{cover2006elements} to show that the probability that the DM-EC causes erasures less than $(1-\alpha) n$ goes to $0$ as $n\rightarrow\infty$.

In particular, we first show that, for $0\leq \lambda < \alpha \leq 1$, we have $\mathsf{C}_{s}^{{\rm{EQ}}}(\alpha)\leq \mathsf{C}_{s}^{{\rm{EQ2}}}(\lambda)$. To do so, we show that every achievable strong secrecy rate for the channel model in Figure \ref{fig:equiv_model} is also achievable for the channel model in Figure \ref{fig:DMC_model}. Fix $\lambda$ such that $0\leq\lambda<\alpha\leq 1$, and let $R_s$ be an achievable strong secrecy rate for the former channel model. Thus, there exists a sequence of $(n,2^{nR_S})$ channel codes $\left\{\setC_n^{\rm{EQ}}\right\}_{n\geq 1}$ such that 
\begin{align}
\limitn\Prob\left(\hat{M}\neq M|\setC_n^{\rm{EQ}}\right)=0,\text{ and }\limitn\max_{S\in\setS}I\left(M;\tbZ_S,\bV|\setC_n^{\rm{EQ}}\right)=0. 
\end{align}

We show that the rate $R_s$ is also an achievable strong secrecy rate for the channel model in Figure \ref{fig:DMC_model} by showing that the sequence of $(n,2^{nR_s})$ codes $\left\{\setC_n^{\rm{EQ}}\right\}_{n\geq 1}$ satisfies the constraints $\limitn\Prob\left(\hat{M}\neq M|\setC_n^{\rm{EQ}}\right)=0$ and $\limitn\max_{S\in\setS}I\left(M;\bZ, \bV|\setC_n^{\rm{EQ}}\right)=0$ for this channel model. 

The main channel in the two models is the same, and hence, the sequence 
of $(n,2^{nR_s})$ codes $\left\{\setC_n^{\rm{EQ}}\right\}_{n\geq 1}$ achieves the reliability constraint for both channel models. Thus, it remains to show that $\left\{\setC_n^{\rm{EQ}}\right\}_{n\geq 1}$ achieves $\limitn\max_{S\in\setS}I\left(M;\bZ,\bV|\setC_n^{\rm{EQ}}\right)=0$.   

Since $\limitn\max_{S\in\setS}I(M;\tbZ_S,\bV)=0$, then for any $\epsilon_0>0$, there exists $n_0\in\mathbb{N}$ such that for all $n\geq n_0$, we have
\begin{align}
\label{eq:converse_16}
\max_{S\in\setS}I\left(M;\tbZ_S,\bV|\setC_n^{\rm{EQ}}\right)=I\left(M;\bV|\setC_n^{\rm{EQ}}\right)+\max_{S\in\setS}I\left(M;\tbZ_S|\bV,\setC_n^{\rm{EQ}}\right)\leq \frac{\epsilon_0}{2}.
\end{align}
Let us define $\tilde{\setZ}=\setX \cup \{?\}$. For every $\bz^n\in\tilde{\setZ}^n$, define 
\begin{align}
\label{eq:converse_17}
&\setN(\bz^n)\triangleq \left\{k\in\llbracket 1,n\rrbracket: z_k=? \right\}\\
\label{eq:converse_18}
&\Theta(\bz^n)=\mathbbm{1}\Big\{|\setN(\bz^n)|\leq \lceil (1-\alpha)n\rceil\Big\}.
\end{align}
That is, $\setN(\bz^n)$ represents the number of erasures in the sequence $\bz^n$, while $\Theta(\bz^n)$ indicates whether the sequence $\bz^n$ has erasures less than or equal to $\lceil (1-\alpha)n\rceil$. 

For simplicity of notation, we drop $\setC_n^{\rm{EQ}}$ from the conditioning in (\ref{eq:converse_16}); it is understood implicitly that the mutual information is calculated with respect to the code $\setC_n^{\rm{EQ}}$. Since $\Theta(\bZ^n)$ is a deterministic function of $\bZ^n$, the Markov chains $M-\bX^n-\bV^n\bZ^n$ and $M\bV^n-\bX^n-\bZ^n$ imply the Markov chains $M-\bX^n-\bV^n\bZ^n\Theta(\bZ^n)$ and $M\bV^n-\bX^n-\bZ^n\Theta(\bZ^n)$. Also, $\Theta(\bZ^n)$ is independent from $\bX^n$. Thus, we have 
\begin{align}
\label{eq:converse_19}
\nonumber p_{\Theta(\bZ)|M\bV}(\theta|m,\bv)&=\sum_{\bx\in\setX^n}p_{\Theta(\bZ)\bX|M\bV}(\theta,\bx|m,\bv)=\sum_{\bx\in\setX^n}p_{\bX|M\bV}(\bx|m,\bv)\;p_{\Theta(\bZ)|\bX M\bV}(\theta|\bx,m,\bv)\\
&=p_{\Theta(\bZ)}(\theta)\sum_{\bx\in\setX^n}p_{\bX|M\bV}(\bx|m,\bv)=p_{\Theta(\bZ)}(\theta)\\
\label{eq:converse_20}
\nonumber p_{\Theta(\bZ)|\bV}(\theta|\bv)&=\sum_{\bx\in\setX^n}p_{\Theta(\bZ)\bX|\bV}(\theta,\bx|\bv)=\sum_{\bx\in\setX^n}p_{\bX|\bV}(\bx|\bv)\;p_{\Theta(\bZ)|\bX\bV}(\theta|\bx,\bv)\\
&=p_{\Theta(\bZ)}(\theta)\sum_{\bx\in\setX^n}p_{\bX|\bV}(\bx|\bv)=p_{\Theta(\bZ)}(\theta).
\end{align}

From (\ref{eq:converse_19}) and (\ref{eq:converse_20}), $M$ and $\Theta(\bZ)$ are conditionally independent given $\bV^n$, and hence,
\begin{align}
\label{eq:converse_21}
I(M;\bZ&|\bV)=I(M;\bZ,\Theta(\bZ)|\bV)\\
\label{eq:converse_22}
&=I(M;\Theta(\bZ)|\bV)+I(M;\bZ|\bV,\Theta(\bZ))\\
\label{eq:converse_23}
&=I(M;\bZ|\bV,\Theta(\bZ))\\
\label{eq:converse_24}
&=\Prob(\Theta(\bZ)=0)I(M;\bZ|\bV,\Theta(\bZ)=0)+\Prob(\Theta(\bZ)=1)I(M;\bZ|\bV,\Theta(\bZ)=1).
\end{align} 
 
We upper bound each term in the right hand side of (\ref{eq:converse_24}). The first term is upper bounded by 
\begin{align}
\label{eq:converse_25}
I(M;\bZ|\bV,\Theta(\bZ)=0)&=I\big(M;\bZ|\bV,\big\{|\setN(\bZ)|>\lceil(1-\alpha)n\rceil\big\}\big)\\
\label{eq:converse_26}
&\leq I\big(M;\bZ|\bV,\big\{|\setN(\bZ)|=\lceil(1-\alpha)n\rceil\big\}\big)\\
\label{eq:converse_27}
&\leq \max_{S\in\setS} I(M;\tbZ_S|\bV).
\end{align}
We also have that
\begin{align}
\label{eq:converse_28}
I(M;\bZ|\bV,\Theta(\bZ)=1)\leq H(\bZ)\leq n\log(|\setX|+1).
\end{align}

Next we upper bound $\Prob(\Theta(\bZ)=1)$. Take $\nu$ such that $\lambda<\nu<\alpha$, and hence, we have $\lceil (1-\alpha)n\rceil \leq (1-\nu)n<(1-\lambda)n$. Let $\Phi_1,\Phi_2,\cdots,\Phi_n$ be a sequence of i.i.d. binary random variables which represents the erasure process of the DM-EC in the model in Figure \ref{fig:DMC_model} ($\Phi_i=1$ if $Z_i=X_i$, and $\Phi_i=0$ if $Z_i=?$), where $\Phi_i$ is distributed according to $Q_{\Phi}={\rm{Bern}}(\lambda)$. Let $Q_{\Phi}^n$ be the $n$-letter distribution of the sequence $\{\Phi_i\}_{i=1}^n$. For each $\xi=\frac{l}{n}$, with $l\in\llbracket\lceil \nu n\rceil,n\rrbracket$, i.e., $\nu\leq \xi<1$, define the distribution $P_{\Phi}^{(\xi)}={\rm{Bern}}(\xi)$, and let $\setP$ be the set of all of these distributions. Let $T(P)$ denote the type class of the distribution $P$, i.e., all possible $n$-length sequences with the type (empirical distribution) $P$ \cite[Section 11.1]{cover2006elements}. Define the set $\setT\triangleq \left\{T(P_{\Phi}^{(\xi)}): (1-\xi)\leq (1-\nu)\right\}$. Using Sanov's theorem \cite[Theorem 11.4.1]{cover2006elements}, we have 
\begin{align}
\label{eq:converse_29}
\Prob(\Theta(\bZ)=1&)=\Prob_{Q_{\Phi}^n}\Big(|\setN(\bZ)|\leq \lceil(1-\alpha)n\rceil \Big)\\
\label{eq:converse_30}
&\leq \Prob_{Q_{\Phi}^n}\Big(|\setN(\bZ)|\leq (1-\nu)n\Big)\\
\label{eq:converse_31}
&=\Prob_{Q_{\Phi}^n}\Big(\big|\left\{k\in\llbracket 1,n\rrbracket: \Phi_k=1\right\}\big|\leq (1-\nu)n\Big)\\
\label{eq:converse_32}
&=\Prob_{Q_{\Phi}^n}(\setT)=\Prob_{Q_{\Phi}^n}(\setP)\leq (n+1)^2\;2^{-n\D(P_{\Phi}^*||Q_{\Phi})},  
\end{align}
where 
\begin{align}
\label{eq:converse_33}
P_{\Phi}^*=\underset{P_{\Phi}^{(\xi)}\in\setP}{\rm{argmin}}\;\D(P_{\Phi}^{(\xi)}||Q_{\Phi})=\underset{\xi:\;\xi\geq \nu}{\rm{argmin}} \left(\xi\log\frac{\xi}{\lambda}+(1-\xi)\frac{1-\xi}{1-\lambda}\right)={\rm{Bern}}(\nu).
\end{align}
Note that $\D(P_{\Phi}^*||Q_{\Phi})>0$ since $\nu\neq \lambda$. 

Using (\ref{eq:converse_28}) and (\ref{eq:converse_32}), the second term in the right hand side of (\ref{eq:converse_24}) is upper bounded by 
\begin{align}
\label{eq:converse_34}
\log(|\setX|+1) n(n+1)^2\;2^{-n\D(P_{\Phi}^*||Q_{\Phi})}\underset{n\rightarrow\infty}\longrightarrow 0.
\end{align}
Thus, for $\epsilon_0>0$, there exists $n_1\in\mathbb{N}$ such that, for all $n\geq n_1$, 
\begin{align}
\label{eq:converse_35}
\Prob(\Theta(\bZ)=1)I(M;\bZ|\bV,\Theta(\bZ)=1)\leq \frac{\epsilon_0}{2}.
\end{align}
Using (\ref{eq:converse_16}), (\ref{eq:converse_24}), (\ref{eq:converse_27}), and (\ref{eq:converse_35}), we have, for sufficiently large $n$, that 
\begin{align}
\label{eq:converse_36}
I\left(M;\bZ,\bV|\setC_n^{\rm{EQ}}\right)&=I\left(M;\bV|\setC_n^{\rm{EQ}}\right)+I\left(M;\bZ|\bV,\setC_{n}^{\rm{EQ}}\right)\\
\label{eq:converse_37}
&\leq I\left(M;\bV|\setC_n^{\rm{EQ}}\right)+\max_{S\in\setS} I\left(M;\tbZ_S|\bV,\setC_n^{\rm{EQ}}\right)+\frac{\epsilon_0}{2}\leq \epsilon_0.
\end{align}
Thus, for $0\leq \lambda <\alpha\leq 1$, we have $\mathsf{C}_{s}^{\rm{EQ}}(\alpha)\leq \mathsf{C}_{s}^{\rm{EQ2}}(\lambda)$. The right hand side of (\ref{eq:upper bound}) is a continuous function of $\alpha$, for $0<\alpha<1$ \cite[Lemma 6]{goldfeld2015semantic}. Thus, by taking $\lambda\rightarrow \alpha$, we have $\mathsf{C}_{s}^{\rm{EQ}}(\alpha)\leq \mathsf{C}_{s}^{\rm{EQ2}}(\alpha)$. Note that for $\alpha=0,1,$ we have $\mathsf{C}_{s}^{\rm{EQ}}(\alpha)= \mathsf{C}_{s}^{\rm{EQ2}}(\alpha)$. Thus, the secrecy capacity of the original model in Figure \ref{fig:sysmodel} is upper bounded by (\ref{eq:upper bound}). This completes the proof for Theorem \ref{thm:Thm1}.

\section{Discussion}\label{Discussion}
In the converse proof for Theorem \ref{thm:Thm1}, we have shown that the strong secrecy capacity $\mathsf{C}_s(\alpha)$ for the new wiretap channel model is equal to the strong secrecy capacity when the wiretapper, in addition to observing $\mu$ transmitted symbols of her choice noiselessly, observes the whole sequence $\bV^n$. This is not surprising because observing noisy versions of the transmitted symbols through the DMC $p_{V|X}$ in the positions where noiseless versions are available does not increase the wiretapper's information about the message. The expression for the strong secrecy capacity of the new wiretap channel model in (\ref{eq:Thm1}) is thus intuitive where $I(U,V)$ represents the secrecy cost due to observing the whole sequence $\bV^n$, and $\alpha I(U;X|V)$ represents the secrecy cost due to observing a fraction $\alpha$ of the transmitted symbols noiselessly, given the wiretapper's knowledge of the $V$ outputs in these positions. Furthermore, the alternative characterization for the strong secrecy capacity of the new wiretap channel model in (\ref{eq:remark_2_1}) is again intuitively pleasing, where the overall secrecy cost is represented by a weighted sum of the secrecy costs due to the noiseless and the noisy observations at the wiretapper, i.e., $\alpha I(U;X)$ and $(1-\alpha) I(U;V)$.

\section{Conclusion}\label{Con}
In this work, we have introduced a new wiretap channel model and derived its strong secrecy capacity. This model generalizes the classical wiretap channel \cite{WTCWyner,CK} to one with a wiretapper who chooses a fixed-length subset of the transmitted codeword symbols to perfectly access, and generalizes the wiretap channel II with a discrete memoryless main channel in \cite{nafea2015wiretap} to one with a wiretapper who observes the output of a noisy channel instead of the erasures. The wiretapper in this model is still passive, yet she is more capable than a classical wiretapper since she can tap a subset of the symbols of her choosing noiselessly, while still receiving the remaining symbols through a channel. Our secrecy capacity result quantifies the secrecy cost of this additional capability of the wiretapper, with respect to the previous wiretap channel models.     

As for future directions, exploring the multi-terminal extensions of this new model is the natural next step, similar to multi-terminal extensions for Wyner's original model, e.g., \cite{tekin2008gaussian,liang2008multiple,lai2008relay,he2013role,ekrem2012capacity2,he2011gaussian}. Additionally, it is of interest to seek new and more powerful wiretapper models against which information theoretic security guarantees can be established.

\appendices
\section{Proof of Lemma \ref{lemma1}}\label{appendix_A}
For $m,c\in \llbracket 1, \tM\rrbracket \times \llbracket 1,\tC\rrbracket$, we have
\begin{align}
\label{eq:appendix_A_1}
P_{MC}(m,c)=\sum_{x\in\setX} p_X(x)\mathbbm{1}\{\setB_1(x)=m\}\mathbbm{1}\{\setB_2(x)=c\}.
\end{align}
We also have that, for all $x\in\setX$, 
\begin{align}
\label{eq:appendix_A_2}
\E_{\setB}\left(\mathbbm{1}\left\{\setB_1(x)=m\right\}\mathbbm{1}\left\{\setB_2(x)=c\right\}\right)=\Prob(\setB_1(x)=m)\Prob(\setB_2(x)=c)=\frac{1}{\tM\tC}. 
\end{align}
Thus, we have $\E_{\setB}({P}_{MC})=\frac{1}{\tM\tC}=p_M^Up_C^U$. For all $m$ and $c$, define the random variables
\begin{align}
\label{eq:appendix_A_3}
&P_1(m,c)=\sum_{x\notin \setD_\gamma} p_X(x)\mathbbm{1}\{\setB_1(x)=m\}\mathbbm{1}\{\setB_2(x)=c\}\\
\label{eq:appendix_A_4}
&P_2(m,c)=\sum_{x\in \setD_\gamma} p_X(x)\mathbbm{1}\{\setB_1(x)=m\}\mathbbm{1}\{\setB_2(x)=c\}.
\end{align}
Note that ${P}_{MC}(m,c)={P}_1(m,c)+{P}_2(m,c)$. Thus, we have 
\begin{align}
\label{eq:appendix_A_5}
&\E_{\setB}\left(\V({P}_{MC},p_M^Up_C^U)\right)=\frac{1}{2}\E_{\setB}\left(\sum_{m,c}\left|P_{MC}(m,c)-\E_{\setB}\left(P_{MC}(m,c)\right)\right|\right)\\
\label{eq:appendix_A_6}
&=\frac{1}{2}\E_{\setB}\left(\sum_{m,c}\left|\sum_{i=1}^2\left(P_i(m,c)-\E_{\setB}\left(P_i(m,c)\right)\right)\right|\right)\\
\label{eq:appendix_A_7}
&\leq \frac{1}{2}\sum_{m,c}\E_{\setB}\left|P_1(m,c)-\E_{\setB}\left(P_1(m,c)\right)\right|+\frac{1}{2}\sum_{m,c}\E_{\setB}\left|P_2(m,c)-\E_{\setB}\left(P_2(m,c)\right)\right|,
\end{align}
where (\ref{eq:appendix_A_7}) follows from the triangle inequality. We now upper bound each term on the right hand side of (\ref{eq:appendix_A_7}). For the first term, we have
\begin{align}
\label{eq:appendix_A_8}
\frac{1}{2}&\sum_{m,c}\E_{\setB}\left|P_1(m,c)-\E_{\setB}\left(P_1(m,c)\right)\right|\leq \sum_{m,c}\E_{\setB}\left(P_{1}(m,c)\right)\\
\label{eq:appendix_A_9}
&=\sum_{m,c} \sum_{x\notin \setD_\gamma}p_X(x)\E_{\setB}\left(\mathbbm{1}\left\{\setB_1(x)=m\right\}\mathbbm{1}\left\{\setB_2(x)=c\right\}\right)\\
\label{eq:appendix_A_10}
&=\sum_{x\notin \setD_\gamma}p_X(x)=\Prob(X\notin\setD_\gamma),
\end{align}
where (\ref{eq:appendix_A_8}) follows from the triangle inequality.

For the second term in the right hand side of (\ref{eq:appendix_A_7}), we have
\begin{align}
\label{eq:appendix_A_11}
\frac{1}{2}&\sum_{m,c}\E_{\setB}\left|P_2(m,c)-\E_{\setB}\left(P_2(m,c)\right)\right|=\frac{1}{2}\sum_{m,c}\E_{\setB}\sqrt{\left(P_2(m,c)-\E_{\setB}\left(P_2(m,c)\right)\right)^2}\\
\label{eq:appendix_A_12}
&\leq \sum_{m,c}\sqrt{\E_{\setB}\left(P_2(m,c)-\E_{\setB}\left(P_2(m,c)\right)\right)^2}\\
\label{eq:appendix_A_13}
&=\sum_{m,c}\sqrt{\Var_{\setB}\left(P_2(m,c)\right)}\leq \frac{1}{2}\sqrt{\frac{\tM\tC}{2^{\gamma}}},
\end{align}
where (\ref{eq:appendix_A_12}) follows from Jensen's inequality and the concavity of square root. The inequality in  (\ref{eq:appendix_A_13}) follows because, for all $m$ and $c$, we have
\begin{align}
\label{eq:appendix_A_14}
\Var_{\setB}\left(P_2(m,c)\right)&=\Var_{\setB}\left(\sum_{x\in \setD_\gamma} p_X(x)\mathbbm{1}\{\setB_1(x)=m\}\mathbbm{1}\{\setB_2(x)=c\}\right)\\
\label{eq:appendix_A_15}
&=\sum_{x\in \setD_\gamma}\Var_{\setB}\Big(p_X(x)\mathbbm{1}\{\setB_1(x)=m\}\mathbbm{1}\{\setB_2(x)=c\}\Big)\\
\label{eq:appendix_A_16}
&\leq \sum_{x\in \setD_\gamma}p^2_X(x)\E_{\setB}\Big(\mathbbm{1}\{\setB_1(x)=m\}\mathbbm{1}\{\setB_2(x)=c\}\Big)\\
\label{eq:appendix_A_17}
&=\frac{1}{\tM\tC}\sum_{x\in\setD_\gamma}p_X^2(x)\\
\label{eq:appendix_A_18}
&\leq \frac{2^{-\gamma}}{\tM\tC}\sum_{x\in\setD_{\gamma}}p_X(x)\leq \frac{2^{-\gamma}}{\tM\tC},
\end{align}
where (\ref{eq:appendix_A_15}) follows since the random variables $\Big\{p_X(x)\mathbbm{1}\{\setB_1(x)=m\}\mathbbm{1}\{\setB_2(x)=c\}\Big\}_{x\in\setD_\gamma}$ are independent due to the structure of the random binning, and (\ref{eq:appendix_A_18}) follows because $p_X(x)\leq 2^{-\gamma}$ for all $x\in\setD_\gamma$. Lemma \ref{lemma1} follows from substituting (\ref{eq:appendix_A_10}) and (\ref{eq:appendix_A_13}) in (\ref{eq:appendix_A_7}).

\section{Proof of Lemma \ref{lemma2}}\label{appendix_B}
We first state the following lemma, which provides a variation of Chernoff bound.
\begin{lemma}\label{lemma6}(A variation on Chernoff bound:)
Let $U_1,U_2,\cdots,U_n$ be a sequence of non-negative independent random variables with respective means $\E(U_i)=\bar{m}_i$. If $U_i\in[0,b]$, for all $i\in\llbracket 1,n\rrbracket$, and $\sum_{i=1}^n\bar{m}_i\leq \bar{m}$, then, for every $ \epsilon\in[0,1]$, we have
\begin{align}
\label{eq:lemma3}
\Prob\left(\sum_{i=1}^n U_i\geq (1+\epsilon)\bar{m}\right)\leq \exp\left(-\epsilon^2\frac{\bar{m}}{3b}\right).
\end{align}
\end{lemma}
\begin{Proof}
The proof is adapted from \cite[Appendix C]{goldfeld2015semantic}. The details are relegated to Appendix \ref{appendix_C}.
\end{Proof}

\subsection{High probability $\setZ$-set:}
For all $S\in\setS$, define the set 
\begin{align}
\label{eq:appendix_B_1}
\setA_S\triangleq\left\{z\in\setZ:\Prob_{p_{X|Z_S}}\left((X,z)\in \setD_{\gamma}^S\right)\geq 1-\delta\right\}. 
\end{align}

Recall that $\Prob_{p_{XZ_S}}\left((X,Z_S)\in\setD_\gamma^S\right)\geq 1-\delta^2$ by assumption. Using Markov inequality, we have 
\begin{align}
\label{eq:appendix_B_2}
\Prob_{p_{Z_S}}(\setA_S^c)&=\Prob_{p_{Z_S}}\left(\Prob_{p_{X|Z_S}}\left((X,Z_S)\notin \setD_{\gamma}^S\right)\geq \delta \right)\\
\label{eq:appendix_B_3}
&\leq \frac{1}{\delta} \E_{p_{Z_S}}\left(\Prob_{p_{X|Z_S}}\left((X,Z_S)\notin \setD_{\gamma}^S\right)\right)\\
\label{eq:appendix_B_4}
&=\frac{1}{\delta}\Prob_{p_{XZ_S}}\left((X,Z_S)\notin\setD_\gamma^S\right)\\
\label{eq:appendix_B_5}
&\leq \frac{\delta^2}{\delta}=\delta.
\end{align}

\subsection{Typical and non-typical events:}
For all $m,c\in\llbracket 1,\tM\rrbracket\times\llbracket 1,\tC\rrbracket$, $z\in\setZ$, and $S\in\setS$, define the random variables  
\begin{align}
\label{eq:appendix_B_6}
&P_1^S(m,c|z)=\sum_{x\in\setX} p_{X|Z_S}(x|z)\mathbbm{1}\{\setB_1(x)=m\}\mathbbm{1}\{\setB_2(x)=c\}\mathbbm{1}\left\{(x,z)\in \setD_{\gamma}^S\right\}\\
\label{eq:appendix_B_7}
&P_2^S(m,c|z)=\sum_{x\in\setX} p_{X|Z_S}(x|z)\mathbbm{1}\{\setB_1(x)=m\}\mathbbm{1}\{\setB_2(x)=c\}\mathbbm{1}\left\{(x,z)\notin \setD_{\gamma}^S\right\}.
\end{align}
Thus, we have, for all $m,c,z,$ and $S$, that 
\begin{align}
\label{eq:appendix_B_8}
P_{MC|Z_S}(m,c|z)&=\sum_{x\in\setX} p_{X|Z_S}(x|z)\mathbbm{1}\{\setB_1(x)=m\}\mathbbm{1}\{\setB_2(x)=c\}\\
\label{eq:appendix_B_9}
&=P_1^S(m,c|z)+P_2^S(m,c|z). 
\end{align}

Note that, for fixed $z\in\setZ$ and $S\in\setS$, each of the the random variables $P_i^S(m,c|z), i=1,2,$ is identically distributed for all $m,c\in\llbracket 1,\tM\rrbracket\times\llbracket 1,\tC\rrbracket$ due to the symmetry in the random binning. We then fix $z\in\setZ$ and $S\in\setS$, and let ${P}_1^S(m,c|z)=\sum_{x\in\setX} U_x(m,c,z,S)$, where  
\begin{align}
\label{eq:appendix_B_10}
U_x(m,c,z,S)=p_{X|Z_S}(x|z)\mathbbm{1}\{\setB_1(x)=m\}\mathbbm{1}\{\setB_2(x)=c\}\mathbbm{1}\left\{(x,z)\in \setD_{\gamma}^S\right\}.
\end{align}  
The random variables $\{U_x(m,c,z,S)\}_{x\in\setX}$ are non-negative and independent, and for all $x\in\setX$, 
\begin{align}
\label{eq:appendix_B_11}
U_x(m,c,z,S)\leq p_{X|Z_S}(x|z)\mathbbm{1}\left\{(x,z)\in \setD_{\gamma}^S\right\}<2^{-\gamma},
\end{align}
where $p_{X|Z_S}(x|z)<2^{-\gamma}$, for all $(x,z)\in \setD_\gamma^S$. Also, we have 
\begin{align}
\label{eq:appendix_B_12}
\nonumber \sum_{x\in\setX}\E_{\setB}&(U_x(m,c,z,S))\\
&=\sum_{x\in\setX} p_{X|Z_S}(x|z)\E_{\setB}\Big(\mathbbm{1}\{\setB_1(x)=m\}\mathbbm{1}\{\setB_2(x)=c\}\Big)\mathbbm{1}\left\{(x,z)\in \setD_{\gamma}^S\right\}\\
\label{eq:appendix_B_13}
&=\frac{1}{\tM\tC} \sum_{x\in\setX} p_{X|Z_S}(x|z)\mathbbm{1}\left\{(x,z)\in \setD_{\gamma}^S\right\}\\
\label{eq:appendix_B_14}
&=\frac{1}{\tM\tC}\;\Prob_{p_{X|Z_S}}\left((X,z)\in\setD_{\gamma}^S\right).
\end{align}

By applying Lemma \ref{lemma6} to the random variables $\{U_x(m,c,z,S)\}_{x\in\setX}$, with $\bar{m}=\frac{\Prob_{p_{X|Z_S}}\left((X,z)\in\setD_{\gamma}^S\right)}{\tM\tC}$ and $b=2^{-\gamma}$, we have, for every $
\epsilon_1\in[0,1]$ and $z\in\setA_S$, that
\begin{align}
\label{eq:appendix_B_15}
\Prob_{\setB}\left(P_1^S(m,c|z)\geq \frac{1+\epsilon_1}{\tM\tC}\right)&\leq \Prob\left(\sum_{x\in\setX} U_x(m,c,z,S) \geq \frac{1+\epsilon_1}{\tM\tC}\;\Prob_{p_{X|Z_S}}\left((X,z)\in\setD_{\gamma}^S\right)\right)\\
\label{eq:appendix_B_16}
&\leq \exp\left(\frac{-\epsilon_1^2\;\Prob_{p_{X|Z_S}}\left((X,z)\in\setD_{\gamma}^S\right) 2^{\gamma}}{3\tM\tC}\right)\\
\label{eq:appendix_B_17}
&\leq \exp\left(\frac{-\epsilon_1^2 (1-\delta) 2^{\gamma}}{3\tM\tC}\right),
\end{align}
where (\ref{eq:appendix_B_15}) follows since $\Prob_{p_{X|Z_S}}\left((X,z)\in\setD_{\gamma}^S\right)\leq 1$, and (\ref{eq:appendix_B_17}) follows because, for all $z\in\setA_S$, we have $\Prob_{p_{X|Z_S}}\left((X,z)\in\setD_{\gamma}^S\right)\geq (1-\delta)$. 

We also have have that,
\begin{align}
\label{eq:appendix_B_18}
&\nonumber \E_{p_{Z_S}}\left(\sum_{m,c} P_2^S(m,c|Z_S)\right)\\
&=\E_{p_{Z_S}}\left(\sum_{x\in\setX} p_{X|Z_S}(x|Z_S)\mathbbm{1}\left\{(x,Z_S)\notin \setD_{\gamma}^S\right\}\sum_{m,c} \mathbbm{1}\{\setB_1(x)=m\}\mathbbm{1}\{\setB_2(x)=c\}\right)\\
\label{eq:appendix_B_19}
&=\sum_{z\in\setZ}p_{Z_S}(z)\sum_{x\in\setX} p_{X|Z_S}(x|z)\mathbbm{1}\left\{(x,z)\notin \setD_{\gamma}^S\right\}=\sum_{(x,z)\notin \setD_{\gamma}^S}p_{XZ_S}(x,z)\\
\label{eq:appendix_B_20}
&=\Prob_{p_{XZ_S}}\left((X,Z_S)\notin \setD_{\gamma}^S \right)\leq \delta^2,
\end{align}
where (\ref{eq:appendix_B_19}) follows since every $x\in\setX$ is assigned to only one pair $(m,c)$, and hence, 
\begin{align}
\label{eq:appendix_B_21}
\sum_{m,c} \mathbbm{1}\{\setB_1(x)=m\}\mathbbm{1}\{\setB_2(x)=c\}=1.
\end{align}

\subsection{Good binning functions:} 
Let $\bold{b}\triangleq(b_1,b_2): \setX\mapsto\llbracket 1,\tM\rrbracket \times \llbracket 1,\tC\rrbracket$ be a realization of the random binning $\setB$. Recall that the random variable $P_1^S(m,c|z)$ is identically distributed  for every $m$ and $c$. We then define the class $\setG$ of binning functions $\bold{b}$ as 
\begin{align}
\label{eq:appendix_B_22}
\setG\triangleq\left\{\bold{b}: P_1^S(m,c|z)<\frac{1+\epsilon_1}{\tM\tC}, \text{ for all } S\in\setS \text{ and all } z\in\setA_S\right\}.
\end{align}
Using the union bound and (\ref{eq:appendix_B_17}), we have that  
\begin{align}
\label{eq:appendix_B_23}
\Prob_{\setB}\left(\setG^c\right)&=\Prob_{\setB}\left(P_1^S(m,c|z)\geq \frac{1+\epsilon_1}{\tM\tC}, \text{ for some } S\in\setS, \text{ or } z\in\setA_S\right)\\
\label{eq:appendix_B_24}
&=\Prob_{\setB}\left(\bigcup_{S\in\setS}\;\bigcup_{z\in\setA_S} {P}_1^S(m,c|z)\geq \frac{1+\epsilon_1}{\tM\tC}\right)\\
\label{eq:appendix_B_25}
&\leq \sum_{S\in\setS}\sum_{z\in\setA_S}\Prob_{\setB}\left(P_1^S(m,c|z)\geq \frac{1+\epsilon_1}{\tM\tC}\right)\\
\label{eq:appendix_B_26}
&\leq \sum_{S\in\setS}|\setA_S|\exp\left(\frac{-\epsilon_1^2 (1-\delta) 2^{\gamma}}{3\tM\tC}\right)\\
\label{eq:appendix_B_27}
&\leq |\setS||\setZ|\exp\left(\frac{-\epsilon_1^2 (1-\delta) 2^{\gamma}}{3\tM\tC}\right).
\end{align} 

Take $\bold{b}$ such that $\bold{b}\in\setG$, and set $M=b_1(X)$ and $C=b_2(X)$. For every $S\in\setS$, we have
\begin{align}
\label{eq:appendix_B_28}
&\D\left(P_{MCZ_S}||p_M^U p_C^U p_{Z_S}\right)=\E_{p_{Z_S}}\left(\D({P}_{MC|Z_S}||p_M^Up_C^U)\right)\\
\label{eq:appendix_B_29}
&=\E_{p_{Z_S}}\left(\sum_{m,c}P_{MC|Z_S}(m,c|Z_S)\log\frac{P_{MC|Z_S}(m,c|Z_S)}{p_M^U p_C^U}\right)\\
\label{eq:appendix_B_30}
&=\E_{p_{Z_S}}\left(\sum_{m,c}\sum_{i=1}^2 {P}_i^S(m,c|Z_S)\log\Big(\tM\tC\sum_{i=1}^2 {P}_i^S(m,c|Z_S)\Big)\right)\\
\label{eq:appendix_B_31}
&=\E_{p_{Z_S}}\left(\sum_{m,c}\sum_{i=1}^2 {P}_i^S(m,c|Z_S)\log\frac{\sum_{i=1}^2 {P}_i^S(m,c|Z_S)}{\frac{1}{\tM\tC} \sum_{i=1}^2\sum_{m,c}{P}_i^S(m,c|Z_S)}\right)\\
\label{eq:appendix_B_32}
&\leq \E_{p_{Z_S}}\left(\sum_{i=1}^2 \sum_{m,c}{P}_i^S(m,c|Z_S)\log \frac{\tM\tC{P}_i^S(m,c|Z_S)}{\sum_{m,c}{P}_i^S(m,c|Z_S)}\right)\\
\label{eq:appendix_B_33}
&\nonumber =\sum_{i=1}^2\E_{p_{Z_S}}\left(\sum_{m,c}{P}_i^S(m,c|Z_S)\log\left(\tM\tC{P}_i^S(m,c|Z_S)\right)\right)\\
&\qquad \qquad +\E_{p_{Z_S}}\left(\sum_{i=1}^2 \sum_{m,c}{P}_i^S(m,c|Z_S)\log \frac{1}{\sum_{m,c}{P}_i^S(m,c|Z_S)}\right)
\end{align}
where (\ref{eq:appendix_B_31}) follows because $\sum_{m,c}\sum_{i=1}^2{P}_i^S(m,c|Z_S)=\sum_{m,c}P_{MC|Z_S}(m,c|z)=1$, and (\ref{eq:appendix_B_32}) follows from the log-sum inequality. 

Now, we upper bound each in (\ref{eq:appendix_B_33}) term. For $\bold{b}\in\setG$ and every $S\in\setS$, we have
\begin{align}
\label{eq:appendix_B_34}
\nonumber &\E_{p_{Z_S}}\left(\sum_{m,c}{P}_1^S(m,c|Z_S)\log\left(\tM\tC{P}_1^S(m,c|Z_S)\right)\right)\\
\nonumber &=\E_{p_{Z_S}}\left(\sum_{m,c}{P}_1^S(m,c|Z_S)\log\left(\tM\tC{P}_1^S(m,c|Z_S)\right)\mathbbm{1}\left\{Z_S\in\setA_S\right\}\right)\\
&\qquad\qquad +\E_{p_{Z_S}}\left(\sum_{m,c}{P}_1^S(m,c|Z_S)\log\left(\tM\tC{P}_1^S(m,c|Z_S)\right)\mathbbm{1}\left\{Z_S\notin\setA_S\right\}\right)\\
\label{eq:appendix_B_34}
&< \log (1+\epsilon_1)+\sum_{x,z}p_{XZ_S}(x,z)\log\left(\tM\tC{P}_1^S(m,c|z)\right)\mathbbm{1}\left\{z\notin\setA_S\right\}\\
\label{eq:appendix_B_35}
&\leq \log (1+\epsilon_1)+\log(\tM\tC)\;\Prob_{p_{Z_S}}(Z_S\notin\setA_S)\\
\label{eq:appendix_B_36}
&\leq \epsilon_1+\delta \log(\tM\tC), 
\end{align}
where (\ref{eq:appendix_B_34}) follows because, for every $\bold{b}\in\setG$ and $S\in\setS$, we have $\tM\tC {P}_1^S(m,c|Z_S)<(1+\epsilon)$ for $Z_S\in\setA_S$ and every $m,c$, and (\ref{eq:appendix_B_36}) follows from (\ref{eq:appendix_B_5}).
 
Using (\ref{eq:appendix_B_20}), we have, for every $S\in\setS$, that
\begin{align}
\label{eq:appendix_B_37}
\nonumber \E_{p_{Z_S}}&\left(\sum_{m,c}{P}_2^S(m,c|Z_S)\log\left(\tM\tC{P}_2^S(m,c|Z_S)\right)\right)\\
&\leq \log(\tM\tC)\;\E_{p_{Z_S}}\left(\sum_{m,c}{P}_2^S(m,c|Z_S)\right)\leq \delta^2\log (\tM\tC).
\end{align}

We also have, for every $S\in\setS$, that 
\begin{align}
\label{eq:appendix_B_38}
\nonumber \E_{p_{Z_S}}&\left(\sum_{i=1}^2 \sum_{m,c}{P}_i^S(m,c|Z_S)\log \frac{1}{\sum_{m,c}{P}_i^S(m,c|Z_S)}\right)\\
&=\E_{p_{Z_S}}\left( H_b\left(\Prob_{p_{X|Z_S}}((X,Z_S)\in\setD_\gamma^S)\right)\right)\\
\label{eq:appendix_B_39}
&\leq H_b\left( \E_{p_{Z_S}}\left(\Prob_{p_{X|Z_S}}((X,Z_S)\in\setD_\gamma^S)\right)\right)\\
\label{eq:appendix_B_40}
&= H_b(\Prob_{p_{XZ_S}}((X,Z_S)\in\setD_{\gamma}^S))\\
\label{eq:appendix_B_41}
&\leq H_b(1-\delta^2)=H_b(\delta^2),
\end{align}
where (\ref{eq:appendix_B_39}) follows from Jensen's inequality and the concavity of $H_b$, and (\ref{eq:appendix_B_41}) follows since $H_b(x)$ is monotonically decreasing in $x\in\left(\frac{1}{2},1\right)$. Equation (\ref{eq:appendix_B_38}) follows since $\sum_{i=1}^2\sum_{m,c}{P}_i^S(m,c|Z_S)=1$, and $\sum_{m,c}{P}_1^S(m,c|Z_S)=\Prob_{p_{X|Z_S}}\left((X,Z_S)\in\setD_\gamma^S\right)$.

By substituting (\ref{eq:appendix_B_36}), \ref{eq:appendix_B_37}), and (\ref{eq:appendix_B_41}) in (\ref{eq:appendix_B_33}), we have, for every $\bold{b}\in\setG$ and $S\in\setS$, that
\begin{align}
\label{eq:appendix_B_42}
\D\left(P_{MCZ_S}||p_M^Up_C^Up_{Z_S}\right)< \epsilon_1+(\delta+\delta^2)\log(\tM\tC)+ H_b(\delta^2)=\tilde{\epsilon}.
\end{align}
Thus, we have 
\begin{align}
\label{eq:appendix_B_43}
\Prob_{\setB}\bigg(\max_{S\in\setS}\;\D\big(P_{MCZ_S}&||p_M^Up_C^Up_{Z_S}\big)\geq \tilde{\epsilon}\bigg)= 1-\Prob_{\setB}\left(\max_{S\in\setS}\;\D\left(P_{MCZ_S}||p_M^Up_C^Up_{Z_S}\right)< \tilde{\epsilon}\right)\\
\label{eq:appendix_B_44}
&=1-\Prob_{\setB}\left(\D\left(P_{MCZ_S}||p_M^Up_C^Up_{Z_S}\right)< \tilde{\epsilon},\; \text{ for all } S\in\setS\right)\\
\label{eq:appendix_B_45}
&\leq 1-\Prob_{\setB}(\setG)=\Prob_\setB(\setG^c)\\
\label{eq:appendix_B_46}
&\leq |\setS||\setZ|\exp\left(\frac{\epsilon_1^2 (1-\delta)2^{\gamma}}{3\tM\tC}\right),
\end{align}
where the inequality in (\ref{eq:appendix_B_45}) follows because (\ref{eq:appendix_B_42}) implies that 
\begin{align}
\Prob_{\setB}\left(\D\left(P_{MCZ_S}||p_M^Up_C^Up_{Z_S}\right)< \tilde{\epsilon},\; \text{ for all } S\in\setS\right)\geq \Prob_{\setB}(\setG).
\end{align}
This completes the proof for Lemma \ref{lemma2}. The analysis in this proof is adapted from \cite[Appendix]{ahlswede1998common}.

\section{Proof of Lemma \ref{lemma6}}\label{appendix_C}
Let $U_1,U_2,\cdots,U_n$ be a sequence of non-negative independent random variables, which satisfy the conditions of the Lemma. For any $\theta>0$, we have
\begin{align}
\label{eq:appendix_C_1}
\Prob\bigg(\sum_{i=1}^n U_i\geq \;&(1+\epsilon)\bar{m}\bigg)=\Prob\left(e^{\theta\sum_{i=1}^n U_i}\geq e^{\theta(1+\epsilon)\bar{m}}\right)\\
\label{eq:appendix_C_2}
&\leq \frac{\E\left(e^{\theta\sum_{i=1}^n U_i}\right)}{e^{\theta(1+\epsilon)\bar{m}}}\\
\label{eq:appendix_C_3}
&=\frac{\prod_{i=1}^n\E\left(e^{\theta U_i}\right)}{e^{\theta(1+\epsilon)\bar{m}}}\\
\label{eq:appendix_C_4}
&\leq \frac{\prod_{i=1}^n\left(1+\frac{e^{\theta b}-1}{b}\E(U_i)\right)}{e^{\theta(1+\epsilon)\bar{m}}}\\
\label{eq:appendix_C_5}
&\leq \frac{\prod_{i=1}^n e^{\frac{e^{\theta b}-1}{b}\bar{m}_i}}{e^{\theta(1+\epsilon)\bar{m}}}\\
\label{eq:appendix_C_6}
&\leq \frac{e^{\frac{e^{\theta b}-1}{b}\bar{m}}}{e^{\theta(1+\epsilon)\bar{m}}}\\
\label{eq:appendix_C_7}
&=\exp\left(-\left[\theta(1+\epsilon)-\frac{e^{\theta b}-1}{b}\right]\bar{m}\right),
\end{align}
where (\ref{eq:appendix_C_2}) follows from Markov's inequality. (\ref{eq:appendix_C_4}) follows because $e^{\theta x}\leq 1+\frac{e^{\theta b}-1}{b}x$ for $x\in[0,b]$, as $e^x$ is a convex function in $x$, (\ref{eq:appendix_C_5}) follows because $1+x\leq e^x$ for all $x\geq 0$, and (\ref{eq:appendix_C_6}) follows because $\sum_{i=1}^n\bar{m}_i\leq \bar{m}.$ 

The value of $\theta$ which maximizes the right hand side of (\ref{eq:appendix_C_7}) is $\theta^*=\frac{1}{b}\ln(1+\epsilon)>0$, for which we have   
\begin{align}
\label{eq:appendix_C_8}
\Prob\left(\sum_{i=1}^n U_i\geq \;(1+\epsilon)\bar{m}\right)\leq \exp\left(-\frac{\bar{m}}{b}\left[(1+\epsilon)\left(\ln(1+\epsilon)-1\right)+1\right]\right).
\end{align}
By considering Taylor's expansion of $x[ln(x)-1]$ around $x=1$, we have, for all $x\geq 1$, that
\begin{align}
\label{eq:appendix_C_9}
x[\ln(x)-1]+1\geq \frac{1}{2} (x-1)^2-\frac{1}{6}(x-1)^3.
\end{align}
We also have, for $x\in[1,2]$, that
\begin{align}
\label{eq:appendix_C_10}
\frac{1}{2} (x-1)^2-\frac{1}{6}(x-1)^3\geq \frac{1}{3}(x-1)^2.
\end{align}
Thus, for all $x\in[1,2]$, we have 
\begin{align}
\label{eq:appendix_C_11}
x[\ln(x)-1]+1\geq  \frac{1}{3}(x-1)^2.
\end{align}
By applying (\ref{eq:appendix_C_11}), with $x=(1+\epsilon)$, to the right hand side of (\ref{eq:appendix_C_8}), we have, for $\epsilon\in[0,1]$, that 
\begin{align}
\label{eq:appendix_C_12}
\Prob\left(\sum_{i=1}^n U_i\geq \;(1+\epsilon)\bar{m}\right)\leq \exp\left(-\frac{\bar{m}}{3b}\epsilon^2\right).
\end{align}

\bibliographystyle{IEEEtran}
\bibliography{MyLib}

\end{document}